\documentclass[10pt,conference,letterpaper]{IEEEtran}
\usepackage{times,amsmath,epsfig}

\usepackage{xspace}
\usepackage{subfigure}
\usepackage{color}
\usepackage{url,hyperref,todonotes,endnotes,amsmath,amssymb,soul}
\usepackage[ruled,vlined,linesnumbered]{algorithm2e}

\newcommand{\mypara}[1]{\vspace{2mm}\noindent\textbf{#1.}}

\definecolor{lightgray}{rgb}{.8,.8,.8}  
\sethlcolor{lightgray} 
\newcommand{\mynote}[1]{\par\noindent\colorbox{lightgray}{\parbox{\linewidth}{#1}}}
\newcommand{\AJ}[1]{\textcolor{brown}{\mynote{AJ:~#1}}}

\newcommand{\KK}[1]{\textcolor{magenta}{\mynote{KK:~#1}}}

\newcommand{\hide}[1]{}

\newcommand{\shufflejoin}{SMJ\xspace}
\newcommand{\memjoin}{BHJ\xspace}
\newcommand{\qrop}{RAQO\xspace}

\title{Query and Resource Optimizations: A Case for Breaking the Wall in Big Data Systems}
\author{%
{Alekh Jindal, Lalitha Viswanathan, Konstantinos Karanasos }%
\vspace{3mm}\\
\fontsize{10}{10}\selectfont\itshape
Microsoft\\
\vspace{5mm}\\
}
\begin{document}
\maketitle
\begin{abstract} 
Modern big data systems run on cloud environments where resources are shared amongst several users and applications.
As a result, declarative user queries in these environments need to be optimized and executed over resources that constantly change and are provisioned on demand for each job.
This requires us to rethink traditional query optimizers designed for systems that run on dedicated resources.
In this paper, we show evidence that the choice of query plans depends heavily on the available resources, and the current practice of choosing query plans before picking the resources could lead to significant performance loss in two popular big data systems, namely Hive and SparkSQL.
Therefore, we make a case for Resource and Query Optimization (or \qrop), i.e., choosing both the query plan and the resource configuration at the same time.
We describe rule-based RAQO and present alternate decisions trees to make resource-aware query planning in Hive and Spark.
We further present cost-based RAQO that integrates resource planning within a query planner, and show techniques to significantly reduce the resource planning overheads.
We evaluate cost-based RAQO using state-of-the-art System R query planner as well as a recently proposed multi-objective query planner.
Our evaluation on TPC-H and randomly generated schemas show that: (i)~we can reduce the resource planning overhead by up to $16x$, and (ii)~RAQO can scale to schemas as large as $100$ table joins as well as clusters as big as $100K$ containers with $100GB$ each.

\end{abstract}

\section{Introduction}
\label{sec:introduction}

Traditional SQL systems (aka databases) pick a query plan to run on a dedicated set of resources or hardware.
Newer cloud computing environments~\cite{aws,azure}, however, offer a shared pool of resources where per-job resources are provisioned dynamically and on demand, via resource managers such as YARN~\cite{yarn} and Mesos~\cite{mesos}.
As a result, SQL-like systems running on these environments, such as Hive~\cite{hive14}, SparkSQL~\cite{sparksql}, SCOPE~\cite{scopeVLDBJ12}, and Redshift~\cite{redshift}, need to pick resources \textit{in addition to} the query plan.
At Microsoft, this is a problem faced by Azure Data Lake~\cite{adl}, which offers analytics as a service, allowing users to simply submit their declarative queries without worrying about optimizing or provisioning resources for those jobs.
The flexibility in resources in cloud environments, coupled with fast provisioning cycles, allows users to easily trade between price and performance, e.g., getting more/faster machines for better performance.
At the same time, the shared nature of resources also means that the requested resources may not be available immediately, e.g., due to a sudden spike in the workload or a change in the cluster capacity, and the applications need to either work with whatever is available or wait for the request to be satisfied.

\begin{figure}[!t]
    \centering
    \includegraphics[width=3in]{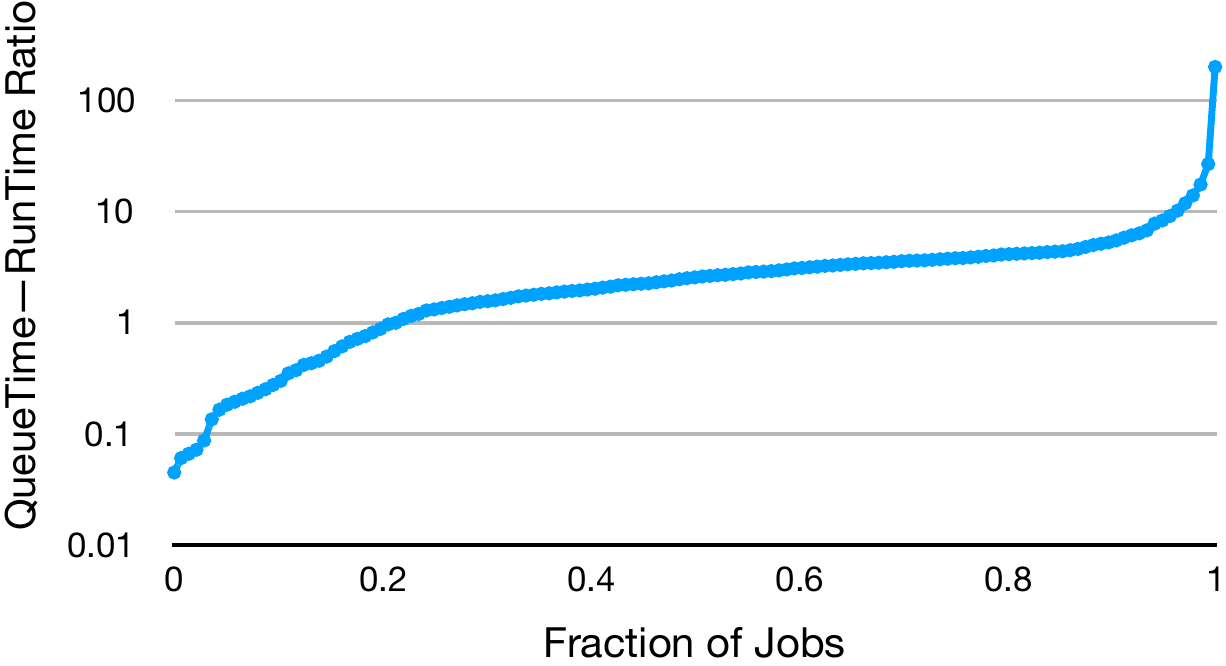}
    \caption{Varying resource availability on Microsoft clusters.}
    \label{fig:resourceWait}
\end{figure}

To illustrate the problem of available versus requested resources, Figure~\ref{fig:resourceWait} shows the cumulative distribution of the ratio of queue-time and execution-time of jobs from one of the business units in production Microsoft clusters. We observe that more than $80\%$ of the jobs spend as much time waiting for resources in the queue as in the actual job execution. More than $20\%$ of the jobs spend at least $4$ times their execution time waiting for resources in the queue.

%

The current practice is to use a \emph{two-step approach} for picking the query and resource plan.
\emph{First}, a query plan is chosen via a query optimizer, e.g., SCOPE optimizer in SCOPE~\cite{scopeVLDBJ12}, Calcite in Hive~\cite{calcite}, or Catalyst in SparkSQL~\cite{sparksql}.
\emph{Second}, the right resource configuration (the \textit{resource plan}) is determined, typically by user guesstimates, simple heuristics, or through a resource optimizer~\cite{perforator,ernest}.

Employing this approach means that \emph{the query and the resource optimizer are not aware of each other}, even when the query and the resource plans heavily depend on each other (e.g., when determining the memory size and the join implementation for join processing in Hive).
Furthermore, the query and resource plans are picked \emph{without considering the current cluster conditions}, which are constantly changing in these environments.
This can result in picking suboptimal query \emph{and} resource plans, thereby leading to significant loss in performance and forcing end users to spend more in order to meet their SLAs.

\begin{figure}[!t]
    \hspace{-0.45cm}
    \subfigure[Hive]{
    \includegraphics[width=0.98\columnwidth/2]{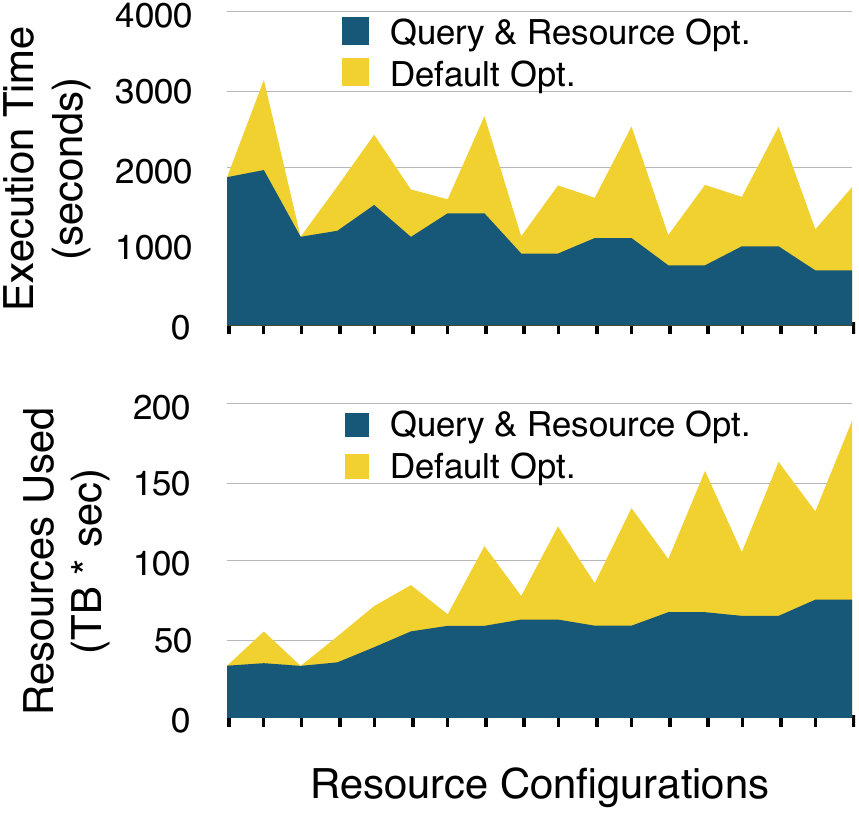}
    \label{fig:gains_hive}
    \vspace{-2mm}
    }
    \hspace{-0.4cm}
    \subfigure[SparkSQL]{
    \includegraphics[width=0.98\columnwidth/2]{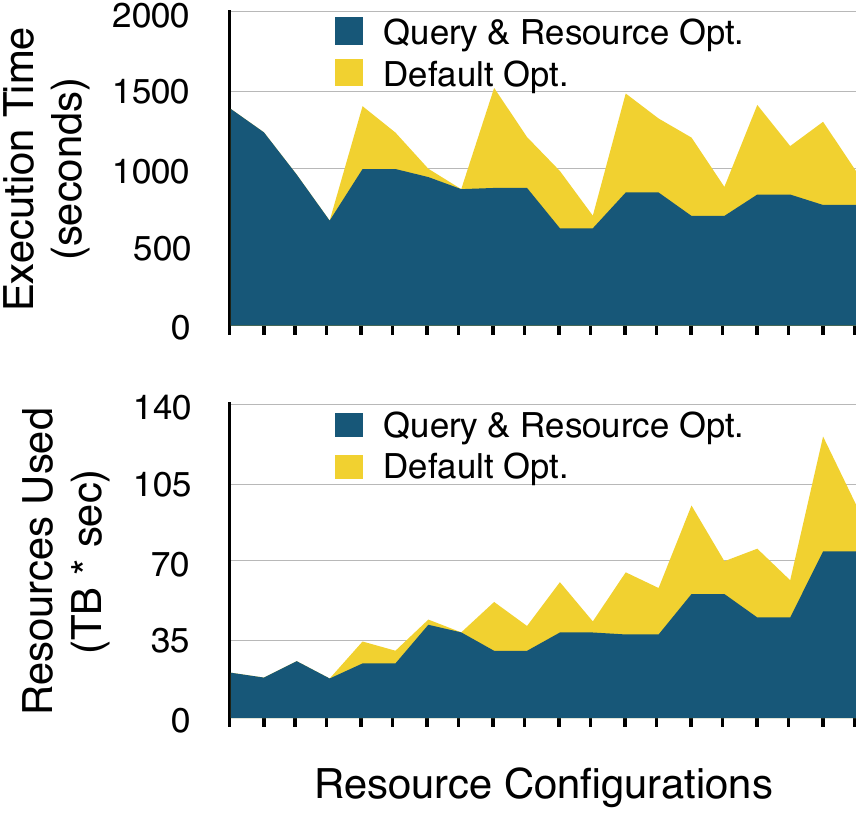}
    \label{fig:gains_spark}
    \vspace{-2mm}
    }    
    \caption{Potential gains of query and resource optimization.}
    \label{fig:potentialGains}
\end{figure}

To motivate the potential gains in case query and resource optimization were aware of each other, we ran a join query on the TPC-H dataset, using both Hive and SparkSQL on a $10$-node YARN cluster, with different join implementations (broadcast and shuffle join) and resource configurations (e.g., memory per container, number of containers).
Figure~\ref{fig:potentialGains} depicts the execution time and the total resources used for each run. 
The total resources are measured as the product of the total memory and the total execution time.
Our results show that the default optimizer picks the optimal plan for very few resource configurations. In particular, the plans chosen by the default optimizer are up to twice slower and twice more resource demanding (which translates to monetary cost) than those chosen by picking the best plan for the given set of resources.

In this work, we propose a redesign of the optimization stack in big data systems. We argue for a more holistic approach, in which the optimizer jointly determines both the query plan and the resource plan, while taking into account the current cluster condition. We term this approach \emph{Resource and Query Optimization} (or \emph{RAQO} for short). We show evidence that the choice of query plans depends heavily on the available resources and the current practice of choosing query plans before picking the resources could lead to significant performance as well monetary loss. 
Motivated from our findings, we describe a novel RAQO architecture and present a rule-based version of it that can be plugged into extensible query optimizers such as Catalyst~\cite{sparksql} and Calcite~\cite{calcite}, and a cost-based version that can be integrated with existing state-of-the-art query planners such as Selinger~\cite{selingerOpt} as well as recently proposed multi-objective query planner~\cite{fastMOQO}.

In summary, our key contributions are as follows:
\begin{itemize}

\item We present a detailed analysis on the cost of ignoring resources for query planning using two popular open source systems, namely Hive and Spark. Our analysis shows that the choice of resources strongly influences the choice of physical operators, the join orderings, as well as the overall monetary costs. (Section~\ref{sec:evidence})

\item We describe the RAQO architecture for a more unified approach to optimizing user queries in big data systems and discuss several use-cases of such an approach. (Section~\ref{sec:architecture})

\item We present a rule-based version of RAQO that can make resource-aware query optimization decisions using a more sophisticated decision tree, which could be easily plugged into popular systems like Hive and Spark. (Section~\ref{sec:dtrees})

\item We present a cost-based version of RAQO that performs resource planning via hill climbing and leverages resource plan caching to further reduce the resource planning overhead. Cost-based RAQO could be integrated with both modern as well as state-of-the-art query planners. (Section~\ref{sec:raqo})

\item We show an evaluation of cost-based RAQO using Selinger and a randomized multi-objective planners. Our results on shows that:
(i)~even with millions of possible resource configurations for TPC-H schema, RAQO nicely combines query and resource planning,
(ii)~RAQO reduces the resource planning overhead by $4x$ via hill climbing and another $4x$ via resource plan caching, and
(iii)~RAQO scales nicely to larger schemas with up to $100$ tables as well to massive cluster sizes with up to $100K$ containers. (Section~\ref{sec:eval})

\item Finally, we discuss our research agenda for the new breed of big data systems that have more holistic query optimization mechanisms. (Section~\ref{sec:research})

\end{itemize}

%
%

Below we first discuss the current state of optimization techniques in big data systems, and then we present each of our contributions.


\section{Background}
\label{sec:background}


More than a decade ago, MapReduce democratized the 
way users could process large amounts of data in parallel~\cite{Dean04,hadoop}. 
Since then, a plethora of systems has been introduced (both from enterprises 
and the academic world, often being released as open-source projects) and a few 
key trends can be observed in their evolution.

First, to alleviate the burden of writing low-level platform-specific code (such as 
MapReduce jobs), most systems started providing users with SQL-like \emph{declarative language abstractions}, e.g., 
Hive~\cite{hive14}, SparkSQL~\cite{sparksql}, and SCOPE~\cite{scopeVLDBJ12}.
Moreover, a richer set of intermediate \emph{dataflow operators} was introduced, 
compared to the initial map and reduce operators~\cite{dryad,spark,tez}. This allows 
different applications (including analytics, machine learning and streaming jobs) 
to be translated to the same intermediate DAG representation\footnote{{\small A DAG consists of vertices (or stages) that correspond to dataflow operators (e.g., map, reduce). Each vertex consists of a set of tasks that can be executed in parallel.}} 
and thus share the 
same execution engine. Consequently, efficiently translating the declarative queries to the underlying DAGs, via a \textit{query optimizer},
becomes crucial.

Second, clusters are now being shared between multiple applications and users in order to make 
better use of their resources. Therefore, the \emph{resource management} layer has been 
abstracted out of the big data systems, forming dedicated resource managers that 
provide an API for multiple systems to use the same cluster. 
Example designs include YARN~\cite{yarn}, Mesos~\cite{mesos}, Apollo~\cite{apollo}, and Borg~\cite{borg}.
Optimizing the resources allocated to each job to improve job completion times and overall cluster resource utilization now becomes important.


The above two trends have also led to two major challenges in big data systems, namely the query and resource optimization.
We discuss these in detail below.

\subsection{Query Optimization}
\label{sec:query-opt}
 
 

Early systems optimized jobs at the dataflow
level~\cite{stubby,stratosphereBlackbox}. For instance, Stubby~\cite{stubby}
optimizes MapReduce workflows, by packing map/reduce functions vertically or
horizontally in order to avoid data shuffling and redundant reads. 
Although useful, these techniques employ black-box optimizations
that are coupled to a particular dataflow engine.

More recent big data systems use a SQL optimizer to translate a given SQL query to an
efficient DAG of operators supported by the underlying dataflow engine. 
For instance, Hive queries get translated to Tez DAGs~\cite{tez}, whereas SparkSQL queries get
translated to Spark DAGs~\cite{spark}. 
Both rule-based~\cite{sparksql} and cost-based~\cite{scopeVLDBJ12,calcite}
approaches have been proposed. Typical optimizations that take place in this
step are filter/projection pushdown, join reordering and choice of operator implementations
(e.g., broadcast or shuffle join, hash- or sort-based aggregation). Some
systems also employ runtime optimizations~\cite{dyno}.
These query optimizations resemble the traditional relational query
optimization~\cite{KossmannSurvey2000}.


 

\subsection{Resource Optimization}
\label{sec:resource-opt}

Along with the translation from SQL query to execution DAGs, the dataflow engine has
to choose the resources to request from the resource manager for
each DAG vertex. We refer to the problem of finding the right resource
configuration as \emph{resource optimization}.
Consider YARN for instance. YARN exposes the cluster resources to applications in the form of
\emph{containers},  which are resource units comprising a fixed amount of
memory and CPU. This is the model  followed by other popular resource managers
too~\cite{mesos,borg}.
Resource optimization with YARN involves determining the \textit{container size}, i.e., the amount of resources per container, the \textit{maximum number of concurrent containers}, i.e., the actual degree of parallelism, and the \textit{total number of containers per DAG vertex}\footnote{{\small In systems like Spark-on-YARN that reuse containers (following the executor model), this is not applicable, although applications still need to determine the number of tasks.}}, i.e., the total tasks per vertex.
%
%

Most common systems rely on the user to provide configurations in the form of parameters (e.g., the
container size in Hive or the degree of parallelism in Spark) or follow simple
heuristics for making such decisions (e.g., Hive determines the number of
reducers based on the intermediate data size). Other systems refine the
resource configurations at runtime after observing actual data
statistics~\cite{scopeVLDBJ12,tez}.
 
Early works in auto-resource tuning studied the problem of provisioning and tuning Hadoop
workloads~\cite{starfish,bazaar}, but their application is limited to MapReduce
systems. Most recently, Ernest~\cite{ernest} and
PerfOrator~\cite{perforator} focused on the resource optimization problem,
relying on executing the job over samples of the input to predict performance and
then pick the right resource configuration. 

Note that all these resource optimizers take as input a \emph{fixed execution
DAG}, which has been produced by a previously performed, \emph{separate query 
optimization step}. Furthermore, these approaches do not take into account the per-operator resources nor the \textit{current}
condition of the cluster.

\section{Cost of Ignoring Resources}
\label{sec:evidence}



In this section, we study the impact of ignoring resources for query planning.
We evaluate the query performance and the monetary costs, both of which the cloud users care about.



\mypara{Setup}
For our analysis we use Apache Hive~\cite{hive14}, a popular open-source system that provides a SQL
interface\footnote{{\small We also did the same experiments with SparkSQL. Due to space limitations, we present detailed analysis on Hive here, and show aggregated results from SparkSQL in \autoref{sec:data-resource-space}.}}.
Hive queries get translated to Tez DAGs, employing Apache Hadoop YARN~\cite{yarn} as the resource manager. 
In particular, we use Hive 2.0.1 on Tez 0.9.0 with 
Hadoop YARN 2.7.2.
We also use the TPC-H dataset~\cite{tpch} with scale factor $100$, and create Hive tables in ORC format.

We conduct our experiments on a cluster of $10$ virtual machines. Each VM has
$4$ cores at $2.2$ GHz, $16$ GB of RAM, and a $3$ TB data drive. VMs are connected
with each other through a $10$ Gbps network. We measure execution times, excluding the overhead of materializing the join output, and report an average of three runs.

We consider the following resource configurations (discussed in \autoref{sec:resource-opt}): (i)~container size, (ii)~maximum number of concurrent containers, and (iii)~total number of containers per DAG vertex. To simplify our analysis, we consider the container sizes in terms of memory (configuration parameter in Hive), but our experiments can naturally be extended to include other resources, such as CPU. 
In our experiments, we change the maximum number of containers by altering the number of VMs of our cluster. In practice, systems like Rayon~\cite{rayon} can be used to determine this parameter. Finally, in the presented results we use a split size of $256$ MB to determine the number of mappers, and enable Hive's feature that automatically determines the number of reducers, since those gave us close to optimal performance.

\subsection{Physical Operators}
\label{sec:single-join}

%


We look at two commonly used join implementations in Hive, namely
\textit{shuffle sort merge join} (\shufflejoin) and \textit{broadcast hash
join} (\memjoin)\footnote{{\small Hive also supports a version of shuffle hash
join. After contacting contributors of Hive, we decided to not
include results for this join, as it is not yet stable enough.}}.
For brevity, here we omit the join implementation details. Note
though that \memjoin, unlike \shufflejoin, broadcasts only the smaller of the
two join relations, avoiding the shuffling of the bigger one. In Hive, 
\memjoin is picked by the optimizer if the smaller relation has size
below a certain threshold (determined through a parameter), with the default
threshold being $10$ MB.

We use the following query: \texttt{select * from orders, lineitem where
o\_orderkey = l\_orderkey}. It is based on
TPC-H query 12, from which we removed the aggregates and additional filters as we want to study only the join behavior.
Note that in our experiments below, we adjusted the smaller \texttt{orders} table size\footnote{{\small To change the \texttt{orders} size, we added a uniform sampling filter on \texttt{o\_orderkey}, which allowed us to select on demand a specific fraction of the table each time.}} proportionally with the resources we had in hand (number and size of VMs).

\subsubsection{Fixed Data, Varying Resources}

\begin{figure}[!t]
    \hspace{-0.2cm}
    \subfigure[Varying Container Size]{
    \includegraphics[width=0.925\columnwidth/2]{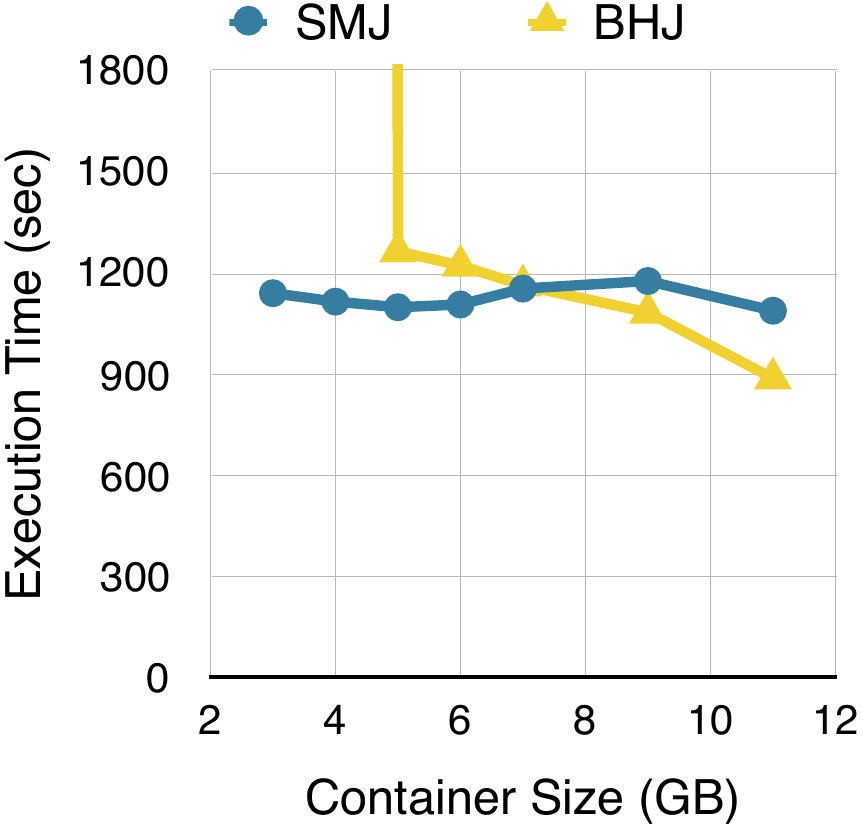}
    \label{fig:varyingResources_hive_containersize}
    }
    \hspace{-0.3cm}
    \subfigure[Varying \#Containers]{
    \includegraphics[width=0.925\columnwidth/2]{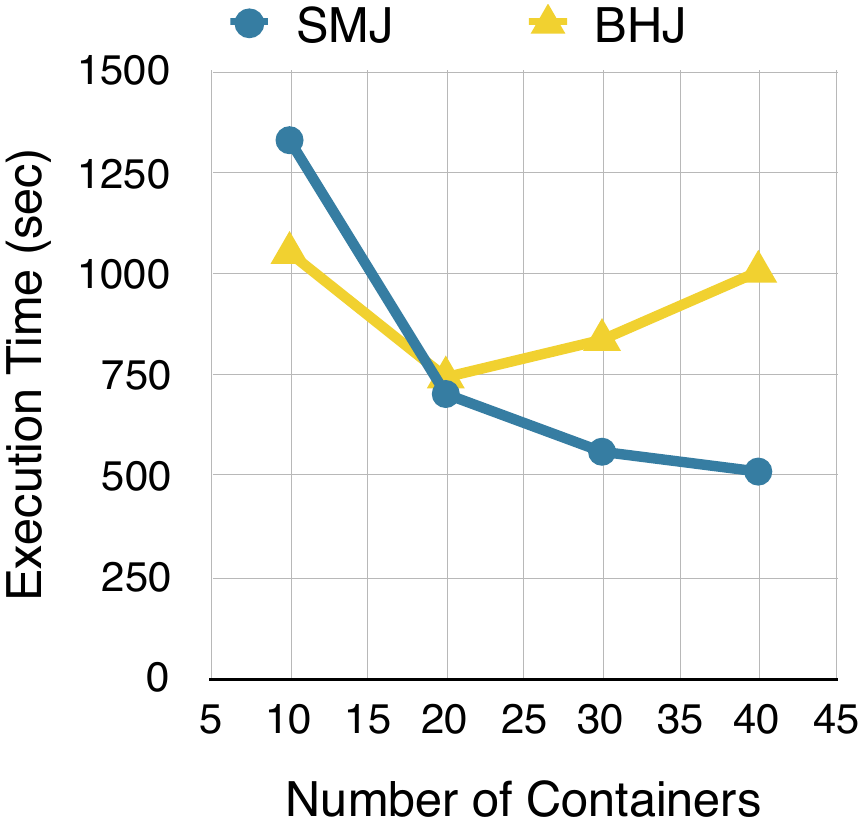}
    \label{fig:varyingResources_hive_numcontainers}
    }    
    \caption{Comparing \memjoin and \shufflejoin over varying resources in Hive.}
    \label{fig:varyingResources_hive}
    \vspace{-0.5cm}
\end{figure}


In this experiment, we present the impact of resource configurations on the execution times of the two join implementations in Hive.
Figure~\ref{fig:varyingResources_hive_containersize} shows the results for our single-join query (with a $5.1$ GB \texttt{orders} table),
using $10$ YARN containers of varying sizes.
Observe that \shufflejoin outperforms \memjoin for container sizes up to $7$ GB, while \memjoin is better for bigger container sizes.
This shows that \memjoin benefits from larger memory, whereas the performance of \shufflejoin remains relatively stable.
Note that below $5$ GB containers, \memjoin is not an option as it runs out of memory with default Hive settings.
Therefore, \emph{the choice between these two implementations depends on the container size}, with the switch point being at $7$ GB for this query.

Figure~\ref{fig:varyingResources_hive_numcontainers} shows the impact of the number of concurrent containers on the execution times, while keeping the size of each container fixed at $3$ GB (using a $3.4$ GB \texttt{orders} table). 
We see that while \memjoin is better than \shufflejoin for less than $20$ containers, \shufflejoin benefits more from increased parallelism and is twice faster than \memjoin for $40$ containers.
Again, we see that \emph{the choice of query plan depends strongly on the number of concurrent containers}, and there is a switch point at $20$ containers in this case.

Thus, \emph{resources do matter when picking a plan} and the current practice in Hive of deciding operator implementations \emph{without looking at the available resources} from YARN could result in significant loss of performance.

\subsubsection{Varying Data and Resources}

We saw above that there is a switch point for choosing operator implementation when varying resources.
The question we will now try to answer is: can these switch points be statically determined and hard-coded into the execution engine or are they dynamic? In the latter case, new query and resource optimizers, i.e., the RAQO architecture, will be required.
To this end, let us see how the switch point changes over varying input sizes.

\begin{figure}[!t]
    \hspace{-0.2cm}
    \subfigure[Container Size]{
    \includegraphics[width=0.925\columnwidth/2]{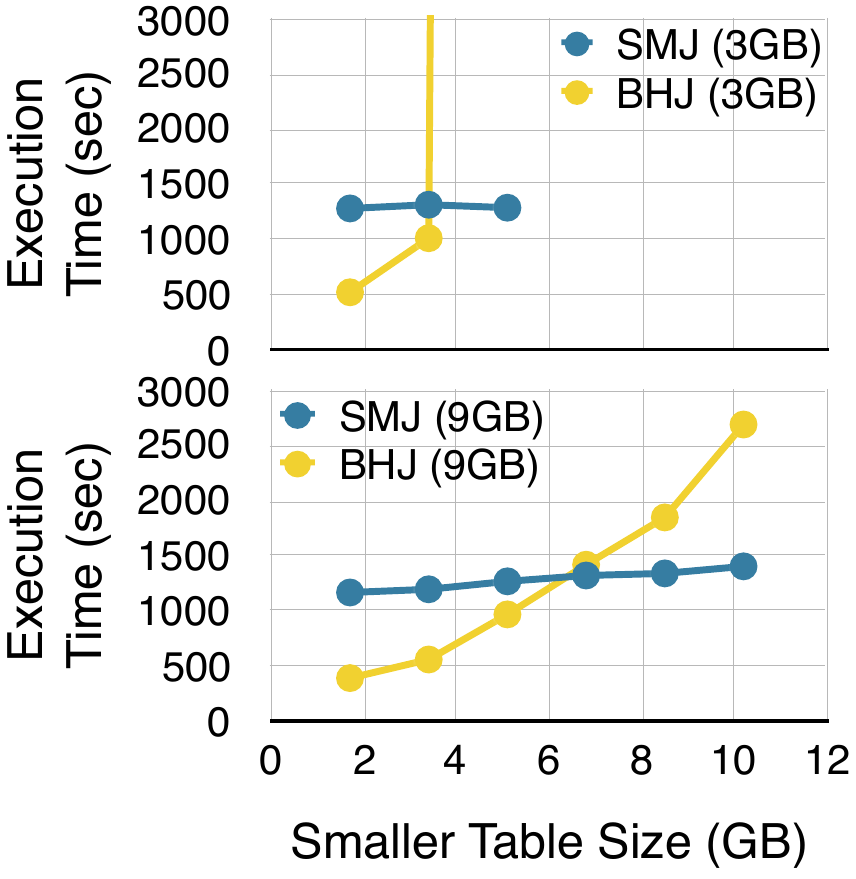}
    \label{fig:hive-varying-data-containersize}
    }
    \hspace{-0.3cm}
    \subfigure[\#Containers]{
    \includegraphics[width=0.925\columnwidth/2]{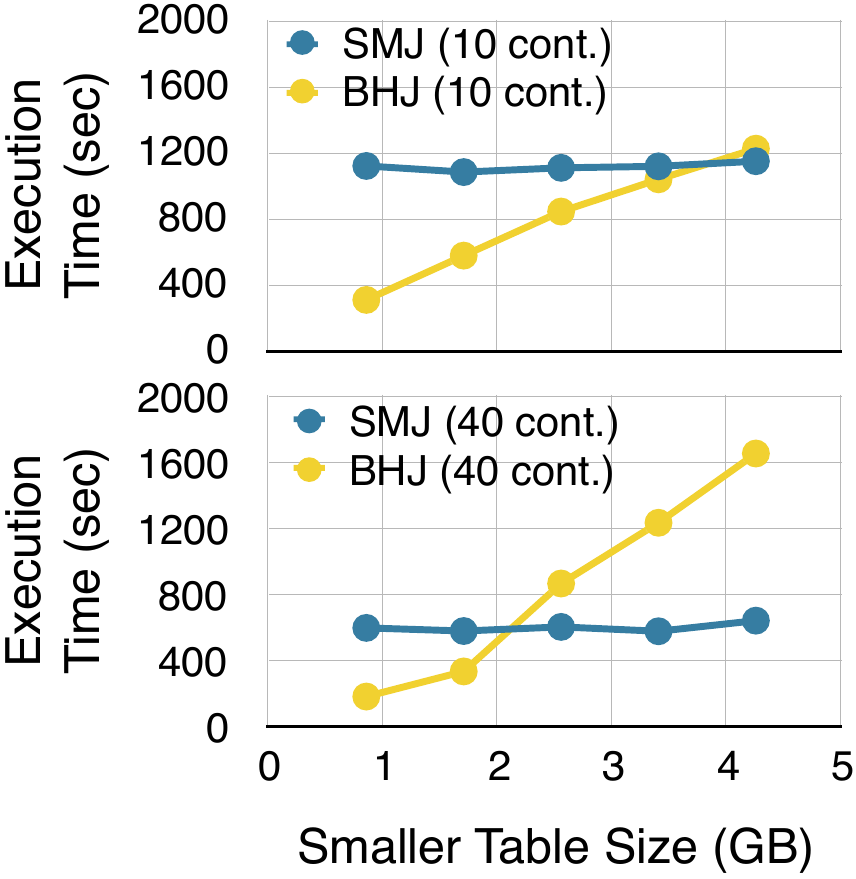}
    \label{fig:hive-varying-data-numcontainers}
    }    
    \caption{Comparing \memjoin and \shufflejoin switch points over varying data size in Hive.}
    \label{fig:hive-varying-data-resources}
\end{figure}

Figure~\ref{fig:hive-varying-data-containersize} shows the execution times when varying the size of the smaller relation (\texttt{orders}) for two different container sizes ($3$ GB and $9$ GB).
We can see that while the switch point between \memjoin and \shufflejoin with $3$ GB containers is at $3.4$ GB of the \texttt{orders}'s size (\memjoin runs out of memory after that), whereas the switch point shifts to $6.4$ GB with $9$ GB containers.
Figure~\ref{fig:hive-varying-data-numcontainers} shows the execution times with different number of concurrent containers, keeping the container size fixed.
Again, we see that the switch point between \memjoin and \shufflejoin shifts from $2.1$ GB of small relation's size with $10$ containers to $3.8$ GB with $40$ containers.
This means that \emph{the switch points are not static and the optimizer has to be aware of both the data statistics (as in traditional query optimizers) and the available resources}.




\subsection{Join Ordering}
\label{sec:multi-join}



\begin{figure}[!t]
    \hspace{-0.1cm}
    \subfigure[Container Size]{
    \includegraphics[width=0.925\columnwidth/2]{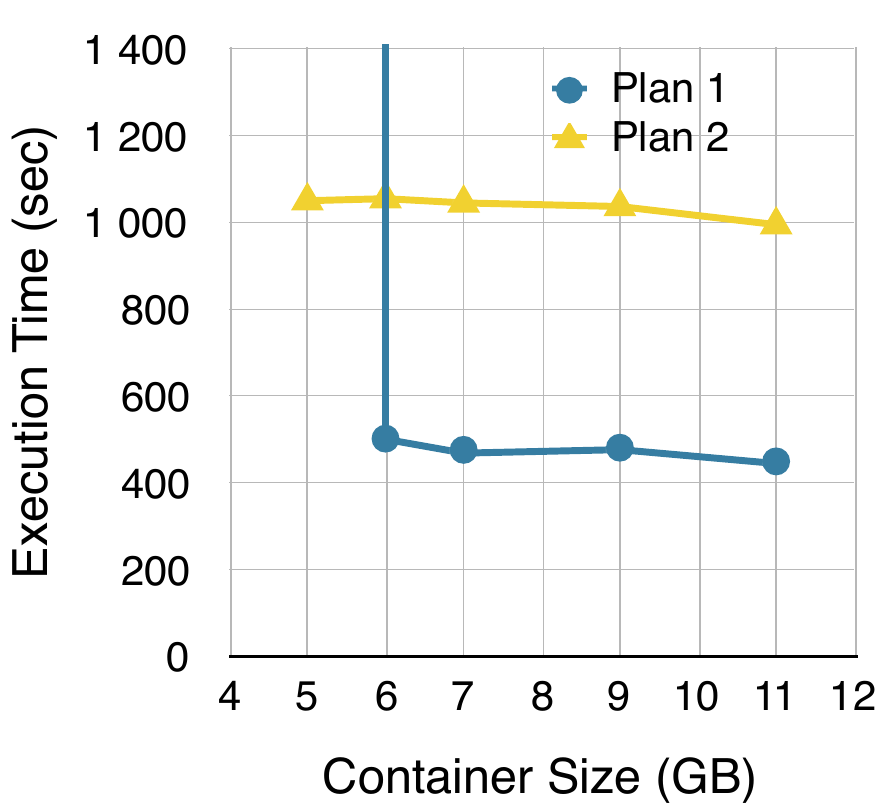}
    \label{fig:hive-jo-containersize}
    }
    \hspace{-0.3cm}
    \subfigure[\#Containers]{
    \includegraphics[width=0.925\columnwidth/2]{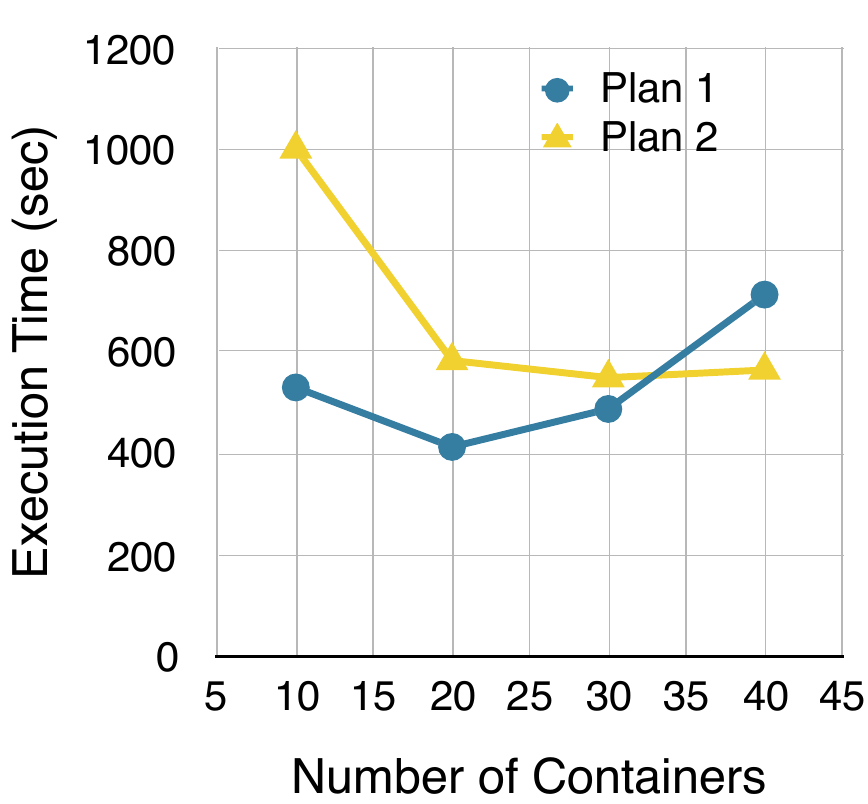}
    \label{fig:hive-jo-numcontainers}
    }
    \caption{Join order decisions in Hive over varying resources.}
    \label{fig:joinordering}
\end{figure}

We now turn to queries comprising multiple operators to study the impact of
resources on different execution plans. To this end,
we use the following two-way join query, which is a simplified version of
TPC-H query 3: \texttt{select * from customer, orders, lineitem where
c\_custkey = o\_custkey and l\_orderkey = o\_orderkey}.  We use part of
\texttt{orders} ($850$ MB in the first experiment below, $425$ MB in the second), so that we can employ more
\memjoin{}s, and make the plan choice more interesting.
We compare two plans on Hive: \\
\emph{Plan 1}: first performs a \memjoin between \texttt{lineitem} and \texttt{orders}, and then a
\memjoin with \texttt{customer}. \\
\emph{Plan 2}: follows a different join order, performing a \memjoin between  \texttt{orders} with \texttt{customer}
and then a \shufflejoin with \texttt{lineitem}.

Figure~\ref{fig:hive-jo-containersize} depicts the execution times for both
plans, using $10$ concurrent containers and different container sizes, while
Figure~\ref{fig:hive-jo-numcontainers} depicts the execution times using $3$
GB containers with varying number of concurrent containers. As shown in the figures, 
container size does not affect execution times significantly and
plan 1 performs better across the board. However, for containers smaller than 
$6$ GB, plan 1 cannot be used as it runs out of memory. On the other hand, the number of
concurrent containers does have an impact on execution times. Interestingly,
when more containers are available, plan 2 starts performing better than plan
1, with $32$ containers being the switch point between the two plans.

\subsection{Monetary Cost}

\begin{figure}[!t]
    \hspace{-0.2cm}
    \subfigure[Varying Container Size]{
    \includegraphics[width=0.925\columnwidth/2]{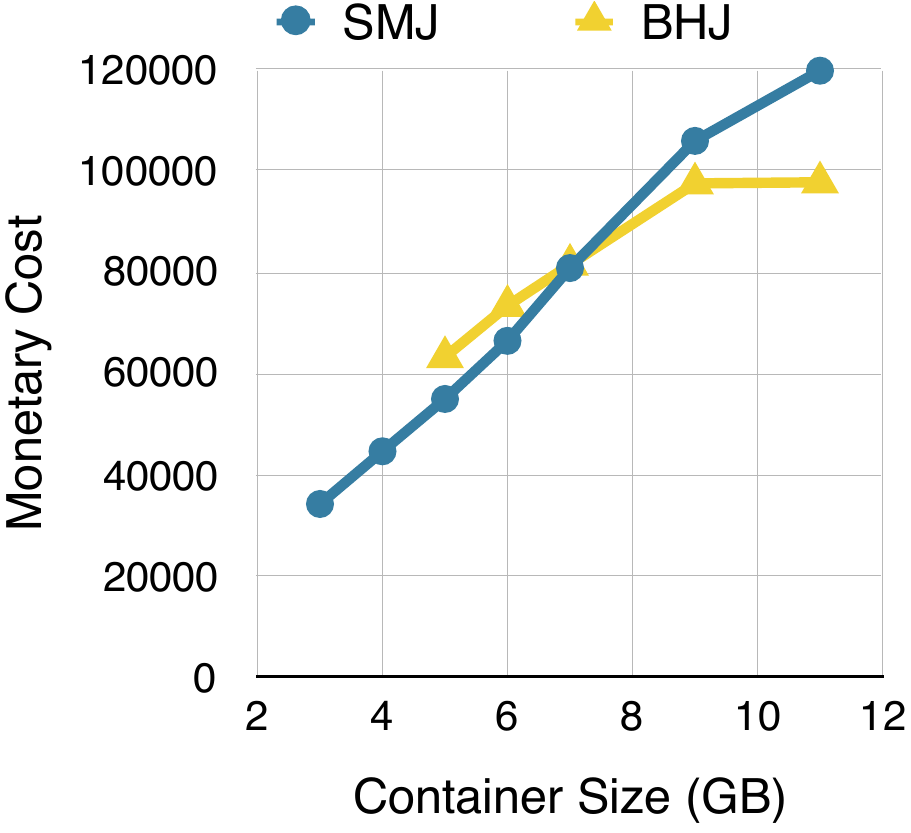}
    \label{fig:varyingResources_hive_containersize-monetary}
    }
    \hspace{-0.3cm}
    \subfigure[Varying \#Containers]{
    \includegraphics[width=0.925\columnwidth/2]{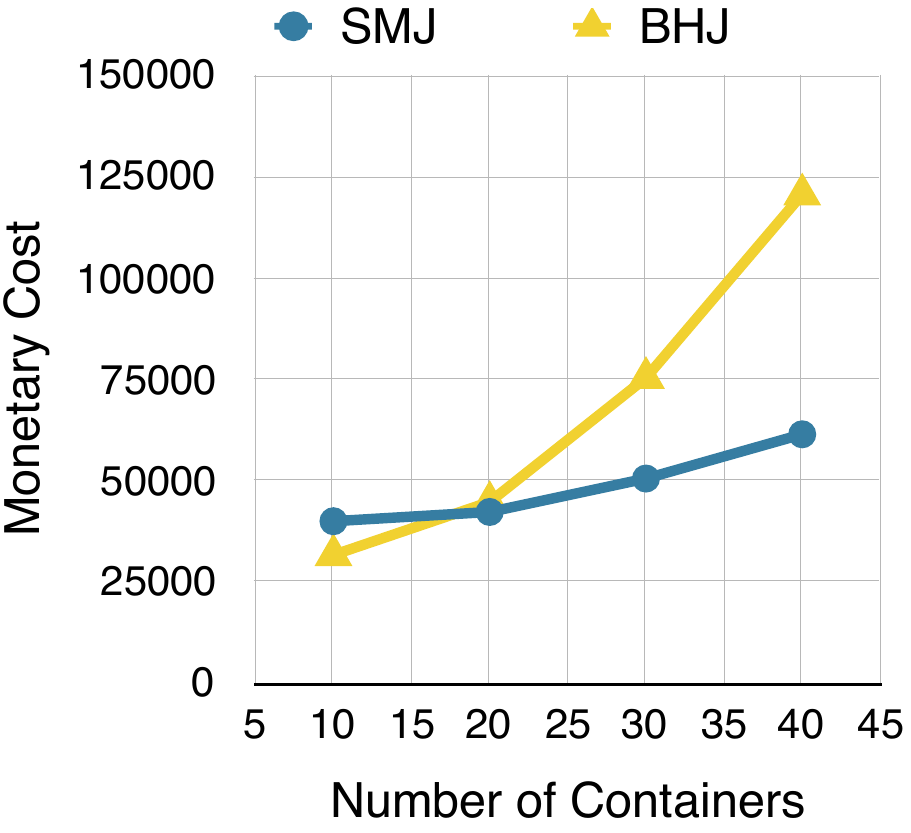}
    \label{fig:varyingResources_hive_numcontainers-monetary}
    }    
    \caption{Comparing monetary cost of \memjoin and \shufflejoin over varying resources in Hive.}
    \label{fig:varyingResources_hive_monetary}
\end{figure}

\begin{figure}[!t]
    \hspace{-0.2cm}
    \subfigure[Container Size]{
    \includegraphics[width=0.925\columnwidth/2]{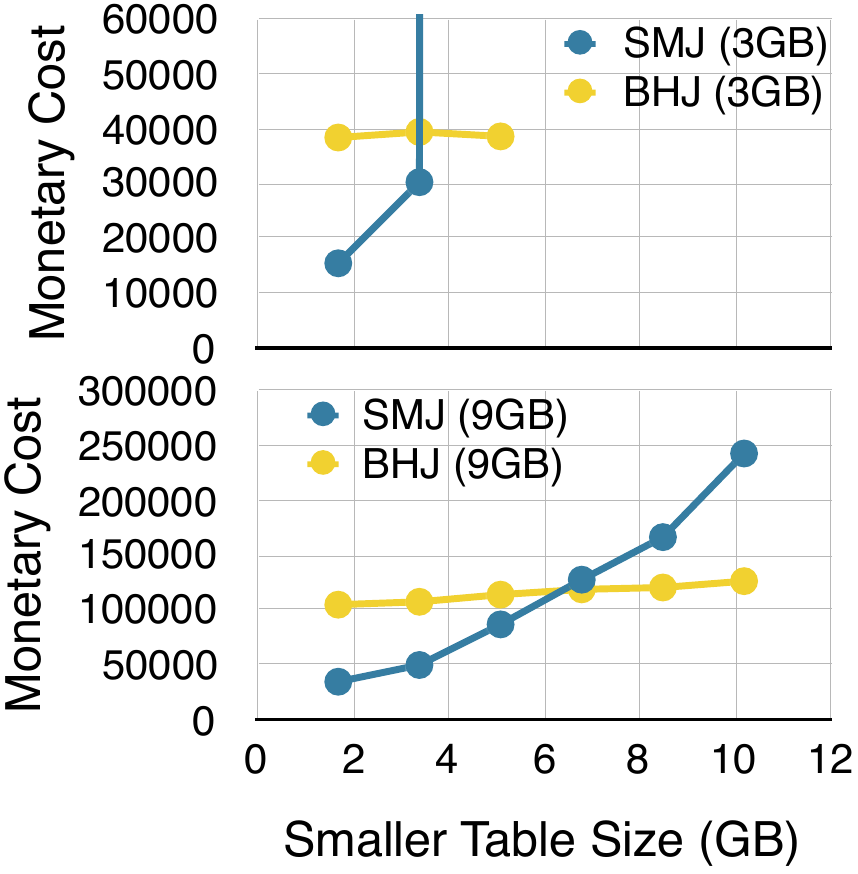}
    \label{fig:hive-varying-data-containersize-monetary}
    }
    \hspace{-0.2cm}
    \subfigure[\#Containers]{
    \includegraphics[width=0.925\columnwidth/2]{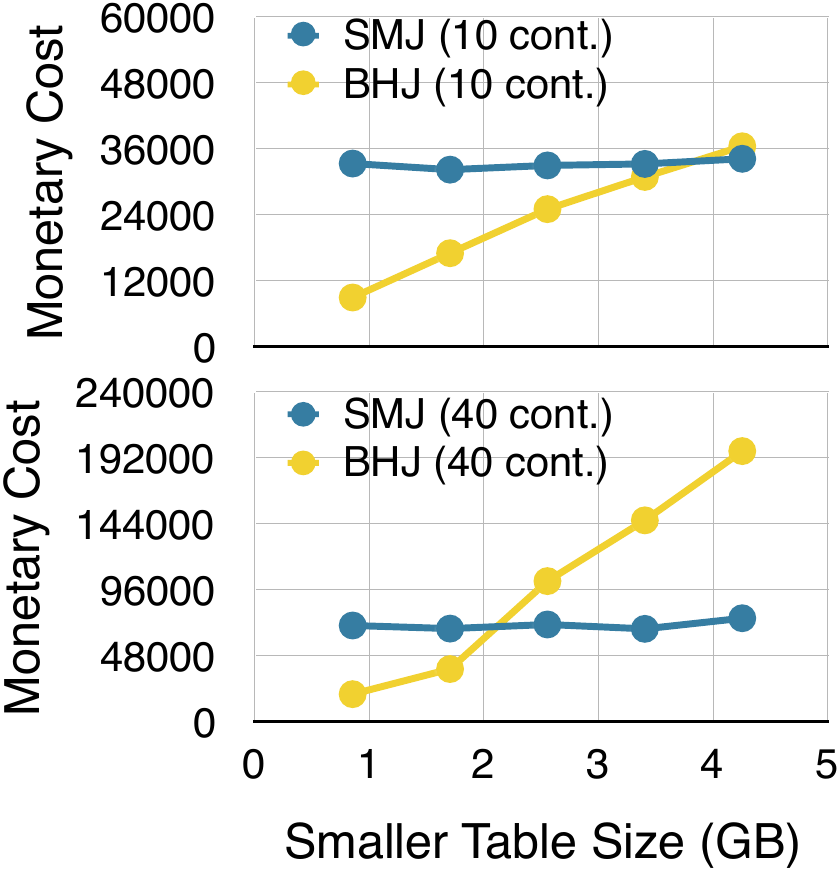}
    \label{fig:hive-varying-data-numcontainers-monetary}
    }    
    \caption{Comparing \memjoin and \shufflejoin monetary switch points over varying data size in Hive.}
    \label{fig:hive-varying-data-resources-monetary}
\end{figure}

In addition to the query performance, cloud users also care about their monetary costs, i.e., the dollar amount they have to pay to run their analytics. 
We consider the recent trend of \textit{serverless} analytics~\cite{adl,athena,bigquery}, where the users only pay for the total container hours consumed by their analytical queries.
The question then is whether resource-aware query planning is important for these monetary costs as well.

Figures~\ref{fig:varyingResources_hive_containersize-monetary} and~\ref{fig:varyingResources_hive_numcontainers-monetary} show the total monetary costs over varying container size 
and number of containers respectively.
Again, we see that either of SMJ and BHJ could be cost effective based on the available resources.
Interestingly, while the switching points remain the same, the absolute values of monetary value change very differently.
While recent works on multi-objective query optimization~\cite{mopqo} could optimize for multiple optimization goals, e.g., performance and monetary costs, these works still lack the notion of resource planning and pick only the query plans that optimizes the set of objectives. Our results suggest that the optimizer needs to carefully pick the resources as well.

Figures~\ref{fig:hive-varying-data-containersize-monetary} and~\ref{fig:hive-varying-data-numcontainers-monetary} show how the switch points between SMJ and BHJ vary over varying data sizes. Again the switch points for most cost effective operator implementation vary both with the available resources as well as the data.
Thus, we see that query planning, without planning for resources, could not only leads to poorer performance but also higher monetary costs.

%
%
%

\hide{
\subsection{Resource Utilization}

\KK{When showing Cosmos data, we need to keep in mind that Cosmos has a single size of containers. Also, about the unutilized resources, Carlo has the over-/under-provisioning data for Cosmos. Given the fixed size of Cosmos containers, I think all that is related to the number of containers only.}

SCOPE/Cosmos have resource boundaries in production; how well are they utilized? 
Show:

-- vertices killed due to exceeding those resource limits

-- un-utilized resources
}

\section{The RAQO Architecture}
\label{sec:architecture}


\begin{figure}[!t]
    \hspace{-0.3cm}
    \subfigure[Current Stack]{
    \includegraphics[height=4.3cm]{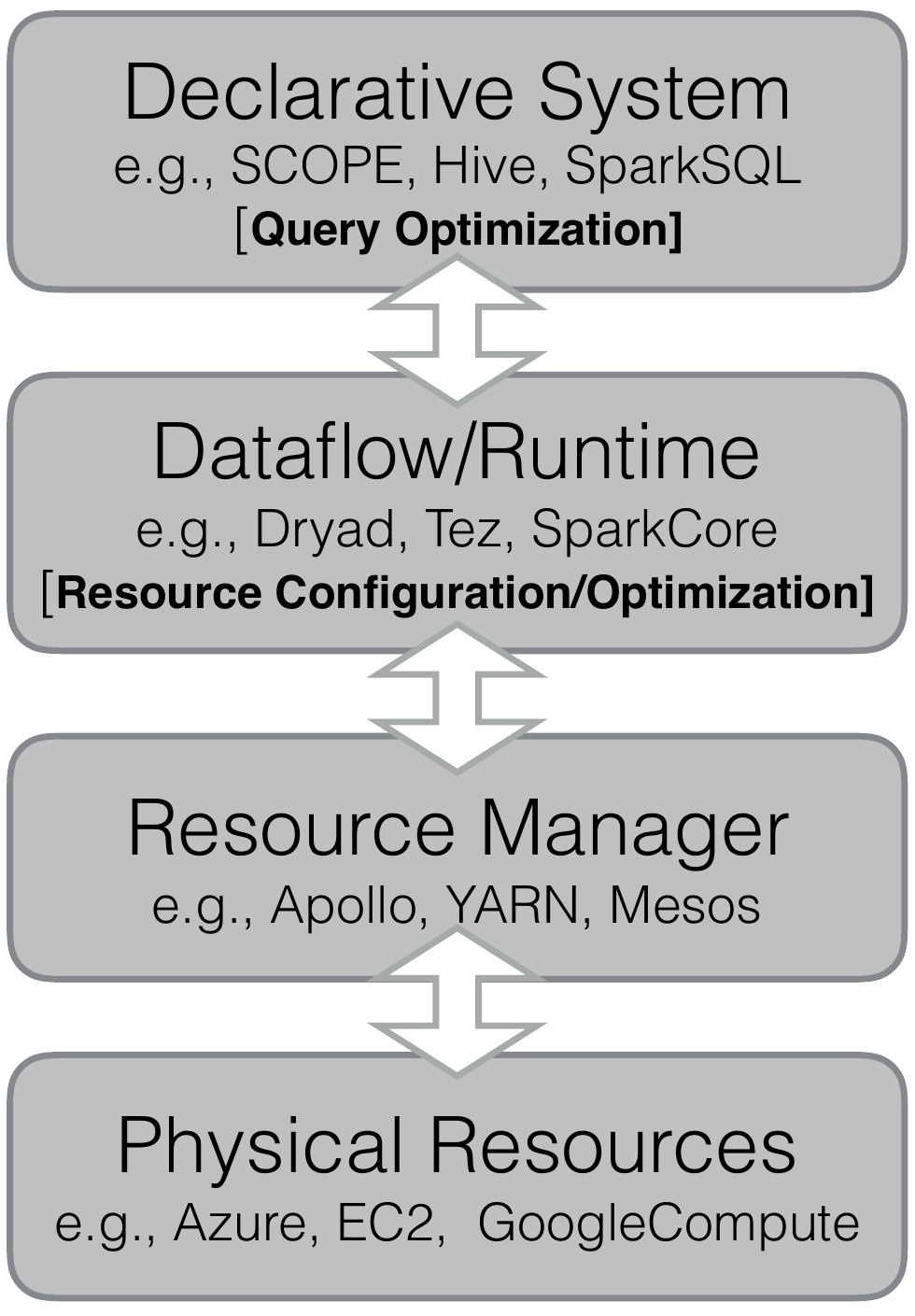}
    \label{fig:stack_before}
    }
    \hspace{-0.2cm}
    \subfigure[RAQO Vision]{
    \includegraphics[height=4.3cm]{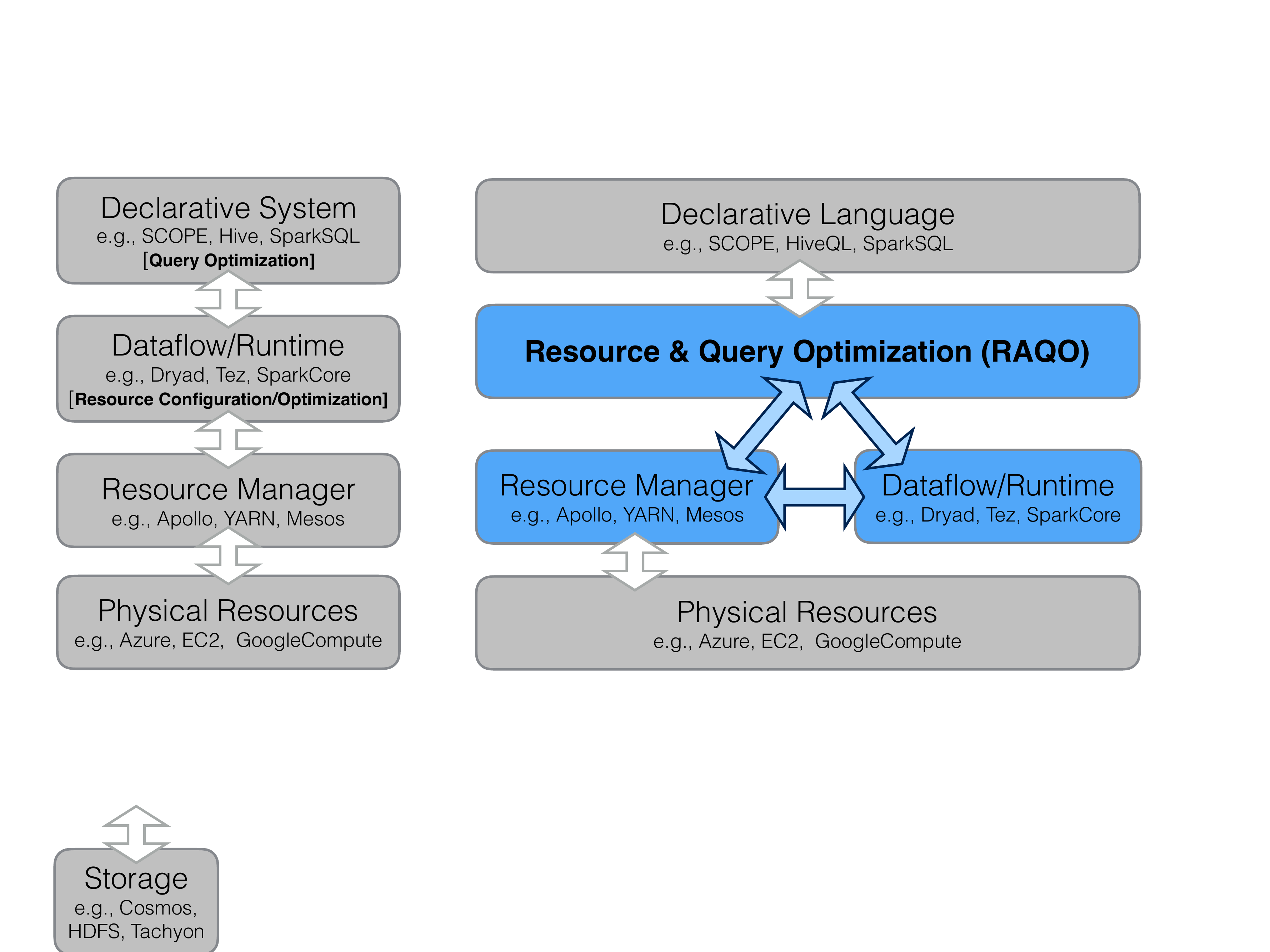}
    \label{fig:stack_new}
    }
    \caption{Big data system stack with an emphasis on Query Optimizer, Resource Optimizer (left), and Query Resource OPtimizer (right).}
    \label{fig:stackComparison}
\end{figure}

The current practice in big data systems is to have query and resource optimization as two separate steps, as shown in Figure~\ref{fig:stack_before}.
First, the query optimizer takes as input the declarative query and produces a physical plan.
Then, the resources are picked for executing the selected physical plan and the resource manager (RM) is invoked for acquiring these resources.
However, we saw in the previous section that ignoring resources leads to bad query plan decisions as well as higher dollar costs.
Therefore, we propose an alternative architecture that combines Resource \textit{and} Query Optimization (RAQO) into a single layer. Figure~\ref{fig:stack_new} illustrates the RAQO architecture.
In this architecture, the \textit{combined optimizer} continuously interacts with the execution runtime as well as the RM.
In its full glory, the optimizer takes as input the declarative query and the current cluster condition (through the RM), and emits a joint query and resource plan, which contains both the operator DAG to be executed by the runtime and the resources to be requested to the RM for each operator in the DAG. However, there could be other variations of RAQO, e.g., emitting only the query plan for a given set of resources, emitting resources for a given query plan, or simply emitting the standard query plan as before. 
If the cluster conditions change until or during the execution of the query, the dataflow/runtime can further adjust the query/resource plan by consulting the optimizer.

The key features of our approach are: (i)~we propose a single holistic optimization for picking both the query plan and the resources, thereby solving two major problems in big data processing at the same time;
(ii)~the new architecture takes into account the condition of the cluster, a dynamically changing feature in shared clusters;
(iii)~there is a tighter coupling between the optimizer and the resource manager, which enables, for instance, to re-optimize query plans in case of any changes in cluster conditions;
(iv)~by emitting a joint query/resource plan, the optimizer can essentially tune the execution time and the monetary cost that the query will yield when run in the cluster/cloud. This is because both the execution time $e$ and the monetary cost $c$ are functions of the query plan $p$ and the resource configuration $r$.

The RAQO architecture enables several interesting use-cases. We enumerate a few below:

\begin{itemize}
\item In case of constrained resources, e.g., with multiple tenants each having their quota, we can pick the best plan for a given resource budget: $r \implies p$.
\item If a user is satisfied with a given performance, e.g., it meets her SLAs, then she can still ask for adjusting the resources to have possibly lower monetary cost:
$p \implies (r,c)$.
\item We can optimize for performance by picking the best query and resource plan combination $(p,r)$. This is useful when there are abundant or even dedicated resources.
\item Instead of resources, we may want to constrain the monetary cost $c$ (a more directly understood metric by the end user). We can then ask the optimizer to adjust the shape of resources (e.g., container size) to produce the best performance for a given price point: $c \implies (p,r)$.
\end{itemize}

Thus, we see that the RAQO architecture opens up new ways of optimizing big data systems, which are more relevant to shared cloud environments and end user needs.


\section{Rule Based RAQO}
\label{sec:dtrees}

\begin{figure}[!t]
    \hspace{-0.2cm}
    \subfigure[Hive]{
    \includegraphics[width=0.9\columnwidth/2]{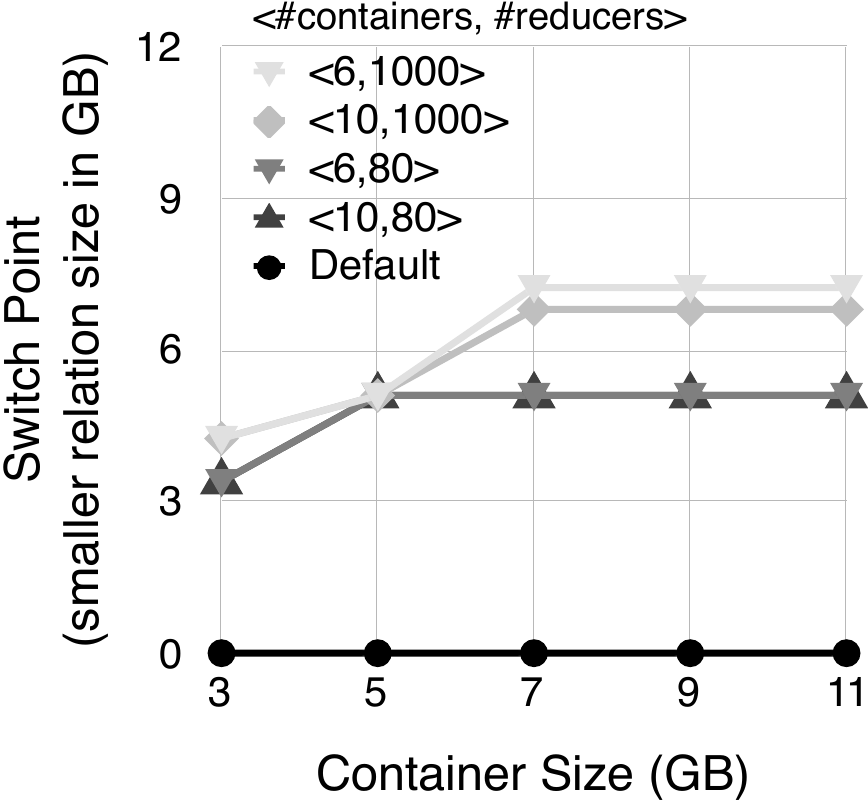}
    \label{fig:hive-qo-space}
    \vspace{-2mm}
    }
    \hspace{-0.3cm}
    \subfigure[Spark]{
    \includegraphics[width=0.95\columnwidth/2]{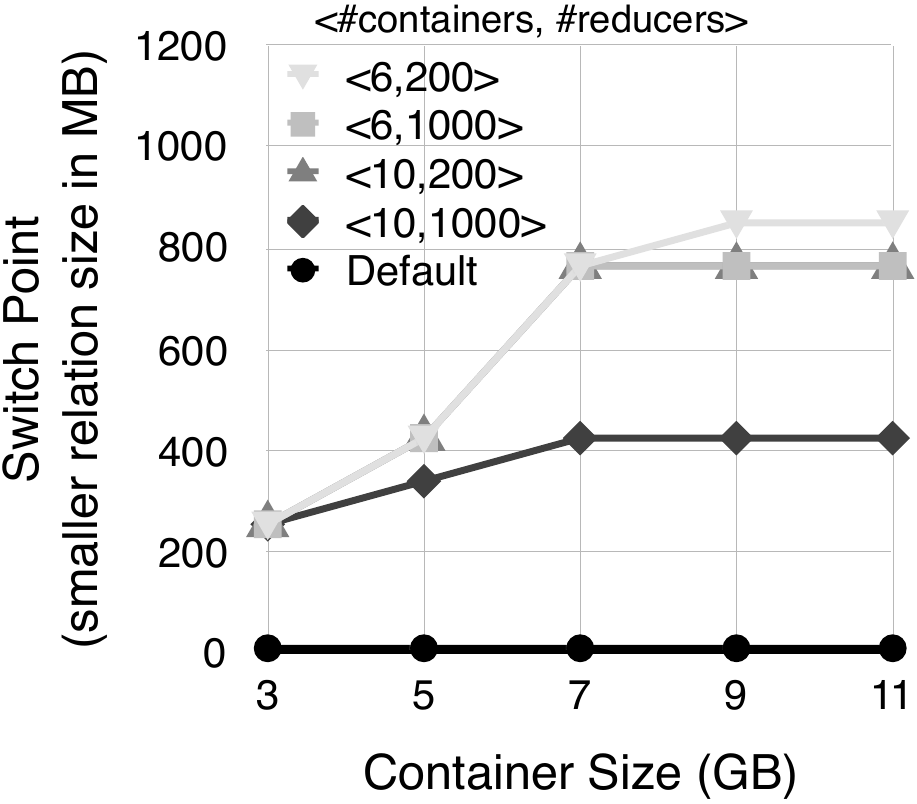}
    \label{fig:spark-qo-space}
    \vspace{-2mm}
    }    
    \caption{The space of \memjoin and \shufflejoin switch points.}
    \vspace{-0.5cm}
    \label{fig:switchpoint-space}
\end{figure}

In this section, we show we could build a rule-based RAQO (Section~\ref{sec:data-resource-space}), and integrate it into existing systems like Hive and Spark (Section~\ref{sec:decision-trees}).

\subsection{The Data-Resource Space}
\label{sec:data-resource-space}

In \autoref{sec:evidence}, we saw the
impact of different resource configurations in isolation, but, in practice, we
need to combine different configurations in the multi-dimensional
data and resource space.

As an example, consider a single-join query, for which we need to pick the
right operator implementation (\memjoin or \shufflejoin) and resources.
Figures~\ref{fig:hive-qo-space} and~\ref{fig:spark-qo-space} show the switch
points in terms of size of the smaller join relation between \memjoin and
\shufflejoin in Hive and Spark\footnote{{\small SparkSQL 1.6.1 running on Hive
tables via the Hive connector.}}, respectively, over different combinations of
container size, number of containers, and number of reducers (we show the
default and optimal number of reducers). For each resource combination, for
small relation sizes within the region below the corresponding curve in the
figure, we suggest choosing a \memjoin, otherwise a \shufflejoin should
be picked.  We also show the default Hive and Spark rules that choose \memjoin
when the small relation is smaller than $10$ MB. Taking into account these
results, along with the execution time of the corresponding runs (not shown in
the figure), allows us to pick the most efficient query/resource plan for the
query.

Key observations from this figure are:  (i)~the optimizer choices change
significantly in this multi-dimensional resource space, (ii)~increasing the
container size helps \memjoin only up to a point, e.g., up to $6$ GB for $10$
containers and $1000$ reducers in Spark, and (iii)~the default optimizer rules
are way off in terms of making the right choices. Finally, the experiment with
Spark also confirms that our observations are not just a Hive artifact, but a more
general big data system phenomenon.

\begin{figure}[!t]
    \centering
    \subfigure[Hive]{
    \includegraphics[width=\columnwidth/2]{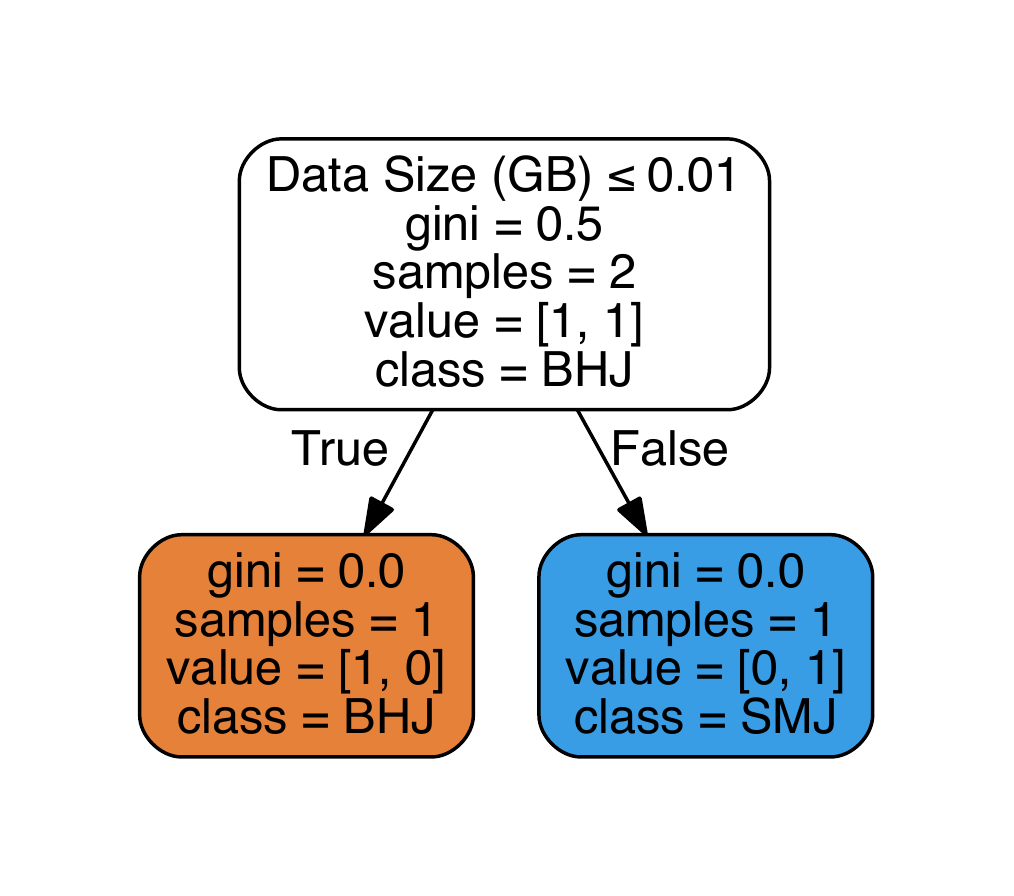}
    \label{fig:hive-default-decision-tree}
    \vspace{-2mm}
    }
    \hspace{-1cm}
    \subfigure[Spark]{
    \includegraphics[width=\columnwidth/2]{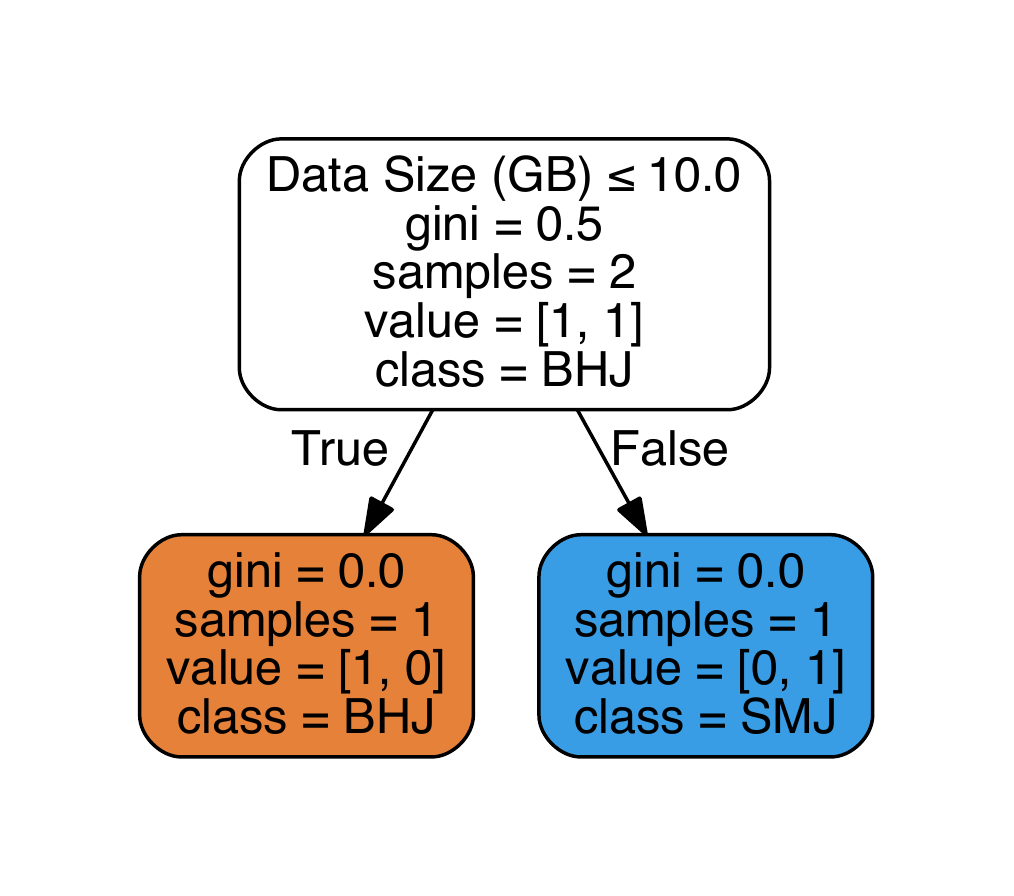}
    \label{fig:spark-default-decision-tree}
    \vspace{-2mm}
    }
    \vspace{-4mm}
    \caption{Default decision trees for join operator implementation in Hive and Spark.}
    \vspace{-0.25cm}
    \label{fig:hive-spark-decision-trees-default}
\end{figure}

\subsection{Decision Trees}
\label{sec:decision-trees}

\begin{figure*}[!t]
    \hspace{-0.3cm}
    \subfigure[Hive RAQO]{
    \includegraphics[width=1.65\columnwidth/2]{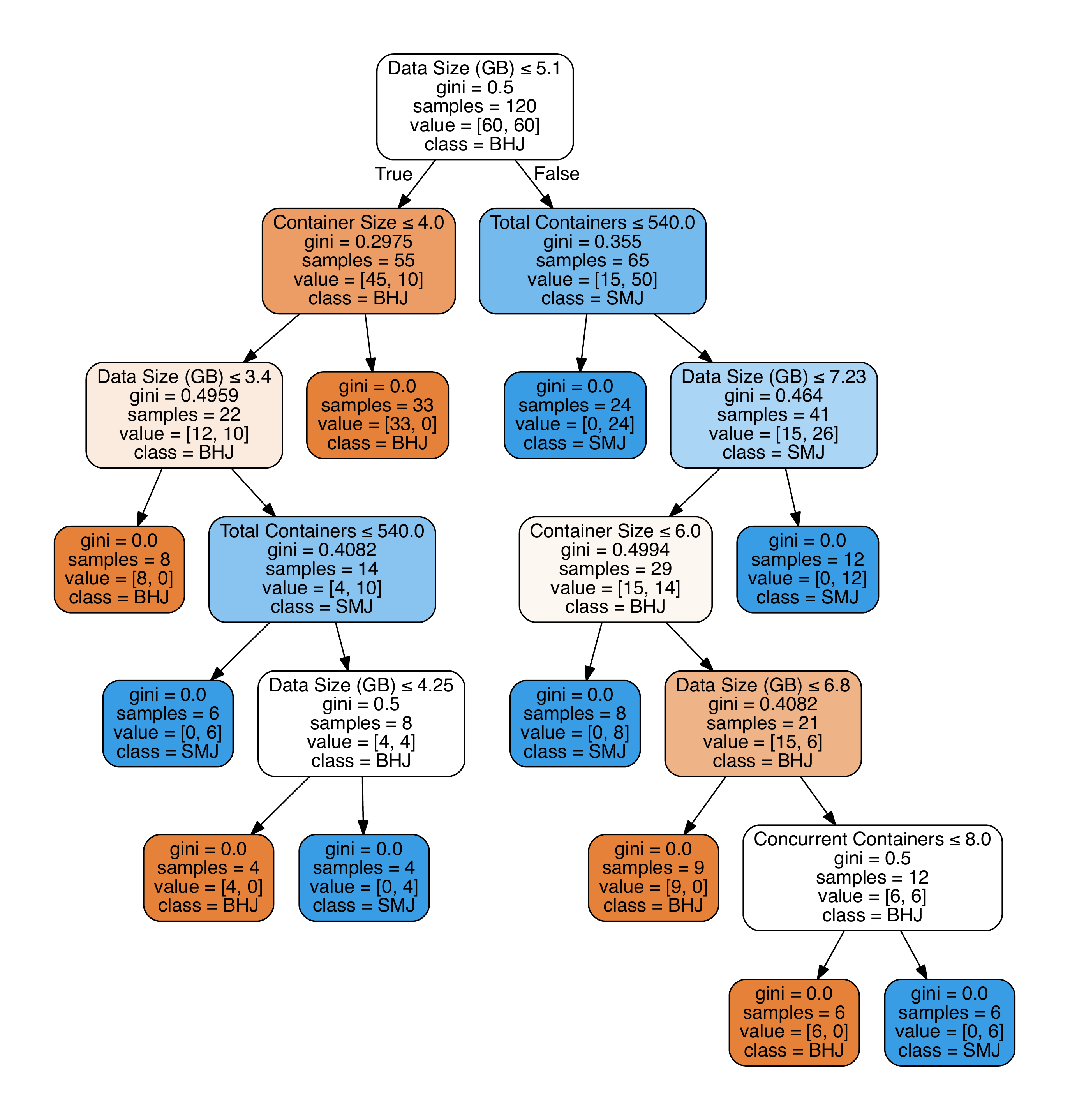}
    \label{fig:hive-qrop-decision-tree}
    }
    \hspace{-1.2cm}
    \subfigure[Spark RAQO]{
    \includegraphics[width=2.5\columnwidth/2]{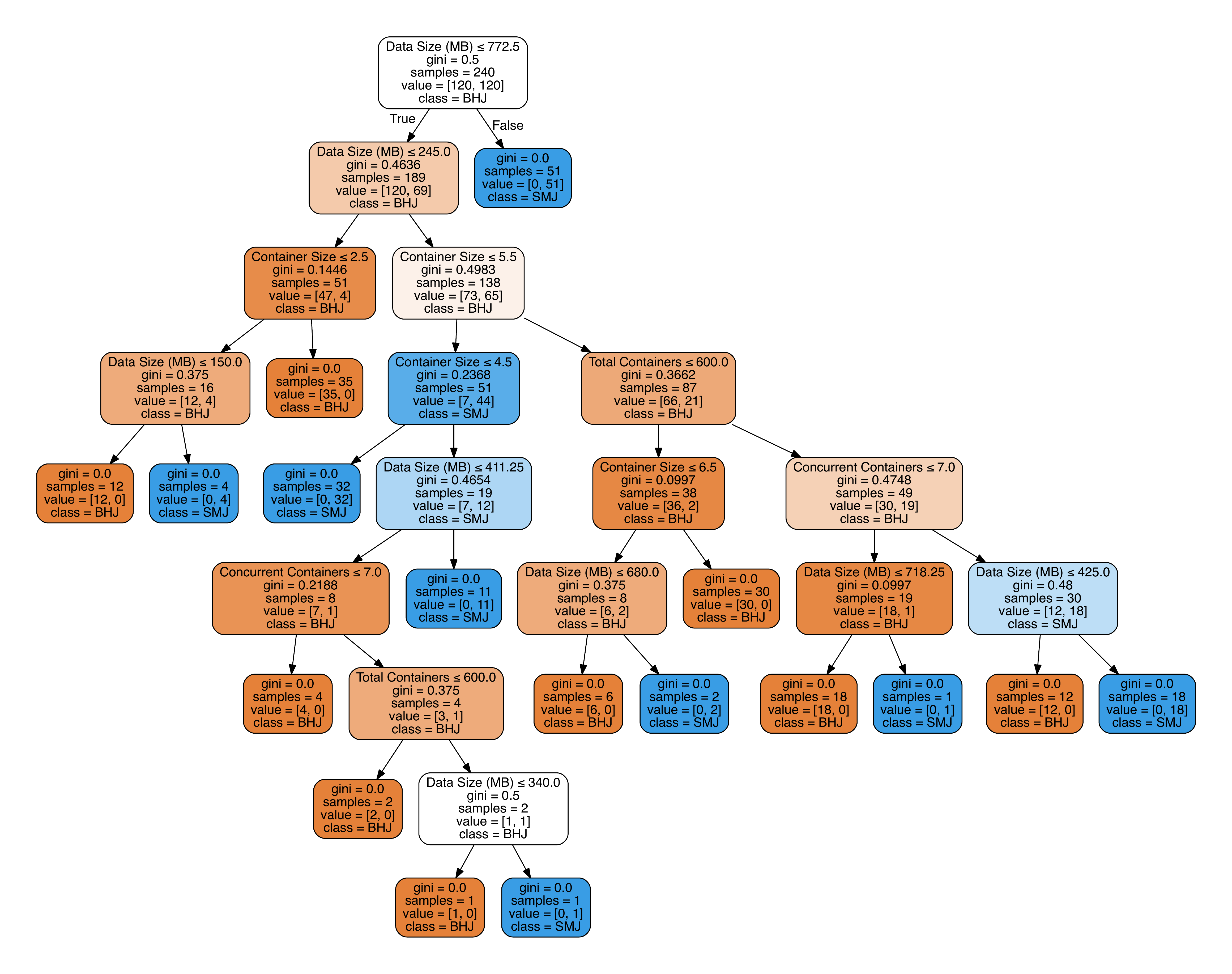}
    \label{fig:spark-qrop-decision-tree}
    }    
    \caption{RAQO decision trees for join operator implementation in Hive and Spark.}
    \label{fig:hive-spark-decision-trees-qro}
\end{figure*}

We can encode our observations from the data-resource space above into a decision tree.
To do this, we ran the decision tree classifier from scikit-learn~\cite{scikit-learn} in python over the switch point results in Figure~\ref{fig:switchpoint-space}, with two target classes namely SMJ and BHJ.
Figure~\ref{fig:hive-spark-decision-trees-default} shows the default Hive/Spark decision trees\footnote{Recall that both Hive and Spark have rule-based join operator selection.} to pick the join operator implementation in Hive/Spark, and Figure~\ref{fig:hive-spark-decision-trees-qro} shows the RAQO decision trees.
The RAQO trees are a bit more complicated, i.e., they have more branching based on not only the data sizes, but also the container sizes and the number of containers.
However, the good thing is that we can simply plug these decisions trees into Hive and Spark in order to make resource aware query planning decisions in those systems.
We still pick the join operator implementations for each join operator in the query DAG independently, however, we use the RAQO decision tree instead.
We traverse the tree using the current cluster conditions (could be the cluster capacity, or could be provided by the resource manager) and the resources available for the query (could be provided by the user).
The leaf of the tree gives the best query plan for those resources.
We can further apply pruning techniques to prune the size of the decision tree~\cite{pessimisticPruning}, however, this is currently not a problem for the set of resources that we have considered in this paper (maximum path length in the RAQO decisions trees is $6$ for Hive and $7$ for Spark.)

Note that building decisions trees as described above is a practical solution since most enterprises that run data analytics have traces of past workload executions (including query plans and resources used), and hence these could be leveraged as training data for the decision trees.

\section{Cost Based RAQO}
\label{sec:raqo}

In this section, we describe our cost-based approach to RAQO.
The goal is to be more general than the rule-based approach, and to be able to estimate costs (and hence pick) from arbitrary query and resource plans.
Below we describe the three steps towards this goal: (i)~coming up with a query and resource cost model, (ii)~techniques for efficiently doing resource planning, and (iii)~integrating resource and query planning into a single optimizer. 


\subsection{Cost Model}
\label{sec:costmodel}

Given the multi-dimensional space of data and resources, we perform a regression analysis to learn the query costs as a function of the input data and resources, i.e., $f(d,r) \rightarrow \mathcal{C}$. Here $\mathcal{C}$ could be a multi-objective cost function. Specifically for our scenario, we trained linear regression models for SMJ and BHJ using smaller input size ($ss$), container size ($cs$), and the number of containers ($nc$) as features. We further augmented the feature set with the following non-linear functions: $ss^2, cs^2, nc^2, \text{ and } (cs\cdot nc)$.
This is to capture non-linear behavior and the interaction between $cs$ and $nc$. The final feature vector is: $[ss,ss^2,cs,cs^2,nc,nc^2,cs\cdot nc]$.
The total cost of a query plan is the sum of costs of all join operators in that plan, i.e., we assume disk-based processing and join operators to be at the shuffle boundaries.
Our regression analysis over the SMJ and BHJ profile runs on Hive yielded the following coefficients:

\begin{small}
\begin{verbatim}
SMJ = [1.62643613e+01, 9.68774888e-01, 
        1.33866542e-02, 1.60639851e-01, 
        -7.82618920e-03, -3.91309460e-01, 
        1.10387975e-01]
BHJ = [1.00739509e+04, -6.72184592e+02, 
       -1.37392901e+01, -1.64871481e+02, 
       2.44721676e-02, 1.22360838e+00, 
       -1.37319484e+02]
\end{verbatim}			
\end{small}

Notice that SMJ has positive coefficients for container size and negative for the number of containers, while it is opposite for BHJ. This makes sense because we saw earlier in Section~\ref{sec:evidence} that SMJ improves more with larger parallelism while BHJ improves more with larger container sizes.

Note that our approach requires profile runs in order to train the cost model. However, this is a one-time investment for each system, e.g., we could tune the system once sufficient workload traces have been collected, or configure the system upfront in case the workload traces are available a priori. Once a model is trained, we could use it to predict costs for arbitrary queries and resource configurations. We could further tune the above cost model by adding more features, trying other training algorithms, and gathering more data points for training. However, that is not the focus of this paper. Our goal here is to demonstrate how query and resource planning could be combined together in a single optimizer. Tuning the cost model itself would be a part of future work.

\subsection{Resource Planning}

We now describe how we pick the resource configurations during optimization.
The total search space for $n$ relations, $a$ operator implementations, $r_p$ possible number of containers, and $r_c$ number of possible container sizes\footnote{Note that both the number of containers as well as the container sizes, typically, have discrete set of possible values.} is given as $n! \cdot (a \cdot r_p \cdot r_c)^n$. Here $n!$ is the number of all possible join orderings for the $n$ relations.
In this paper, we assume that each operator in each of the candidate query plans can make independent decisions on the resource configurations it wants to use.
This is a reasonable assumption because:
(i)~we consider operators across shuffle boundaries (mainly joins), and hence they could have resource configurations allocated independently; other operators could be simply pipelined on same machines,
(ii)~we can extract better performance (and lower costs) by considering all possible resource configurations for each join stage independently, and
(iii)~modern cloud environments increasingly provide instant scaling of resources.
This assumption reduces the search space to $n! \cdot a \cdot n \cdot r_p \cdot r_c$.

\subsubsection{Brute force approach} 

The brute force approach to resource planning would perform an exhaustive search of all possible resource configurations to find the best one.
Essentially, this means running the following set of nested loops and returning the plan with the cheapest cost:
\begin{small}
\begin{verbatim}
For every p in candidate Plans P
  For every s in Subplans of p
    For every o in operator Implementations O
      For every r in Resource Configurations R
        Measure the cost of p with o and r
\end{verbatim}
\end{small}
\vspace{-2mm}
The brute force resource planning quickly grows with large number of resource configurations or larger schemas (see Section~\ref{sec:qroEvalOpt} for more details). 
To address this, below we present our hill climbing method for resource planning. Thereafter, we describe a resource plan cache scheme to further reduce the resource planning costs.

\subsubsection{Hill climbing method} 

We now describe our hill climbing method for resource planning.
Given that the users want to minimize the resources used in modern cloud infrastructures, our idea is to start from the smallest resource configuration and then climb the resource configuration hill until no more improvements in the cost metric could be achieved, i.e., we reach a local optima.

\begin{algorithm}[!t] 
\DontPrintSemicolon
\SetKwInOut{Input}{Input}
\SetKwInOut{Output}{Output}
\SetKwFunction{GetDiscreteSteps}{GetDiscreteSteps}
\SetKwFunction{GetJoinCost}{GetCost}
\SetKwRepeat{Do}{do}{while}
\SetKw{and}{and}

\Input{CostModel $m$, Subplan $p$, Resource $start$, ClusterConditions $clusterCond$}
\Output{the resources to use for executing $p$}
\BlankLine

stepSize = \GetDiscreteSteps{clusterCond}\;
candidate = [-1,1]\;
currRes = start\;

\While{true} {
	currCost = \GetJoinCost{m,plan.d,currRes}\;
	bestCost = currCost\;
	\ForEach{$i \in [0,resourceDims)$} {
		best = -1\;				
		\ForEach{$j \in [0,candidate.length)$} {
			iVal = stepSize[i] * candidate[j]\;
			\If{currRes[i] + iVal $<=$ cluster.max[i] \and currRes[i] + iVal $>=$ cluster.min[i]} {
						currRes[i] += iVal\;
						temp = \GetJoinCost{m,plan.d,currRes}\;
						currRes[i] -= iVal\;
						\If{temp $<$ bestCost} {
							bestCost = temp\;
							best = j\;							
						}
			}
		}
		\If{best $!= -1$} {
			currRes[i] += stepSize[i] * candidate[best]\;
		}
	}			
	\If{bestCost $>=$ currCost} {
		\tcp{return current node since no better neighbors exist}
		\Return currRes\;
	}
}
\caption{HillClimbResourcePlanning} \label{hcResourcePlanning} \end{algorithm}

Algorithm~\ref{hcResourcePlanning} shows the pseudocode for resource planning via hill climbing.
The input to the algorithm is a cost model, as described in Section~\ref{sec:costmodel}, a subplan (a single join operator for now) for which the resources need to be planned, the starting resource configuration for the hill climb (typically minimum possible set of resources), and the current cluster conditions (mainly providing the minimum and maximum cluster resources available currently). The algorithm starts with gathering the hill climb step sizes along all resource dimensions, and initializing the candidate steps to be considered, namely one forward and one backward step (Lines 1--2). In each iteration, the algorithm computes the current cost and then considers stepping along each of the resource dimensions (Lines 5--19). For each resource dimension, we apply each of the candidate steps (given that it does not exceed the cluster conditions), compute the corresponding cost, and backtrack (Lines 10--14). Only when a step produces lower cost than the current one, we reapply the step and set that as the current best cost (Lines 15--19). The algorithm returns when no better resource configuration could be found, i.e., there is no cost improvement from any of the candidate steps (Lines 20--21).

The hill climb resource planning described above not only terminates sooner at a local optimal, but it also climbs the resource space greedily along whichever dimensions provide successive improvements. As a result, it allows us to plan resources with much lower overhead to query planning (we will demonstrate this in Section~\ref{sec:eval}).

\subsubsection{Resource plan cache}

We now present a scheme for caching and reusing resource plans in order to further reduce the resource planning overhead.
Our key insight is that for the same cost model and sub-plan (e.g., join operation), same (or similar) data characteristics, e.g., data size, will require same (or similar) resource configuration.
This means that a resource configuration computed for one join operator in a query tree could be applied to another join operator in the same tree in case they have similar data characteristics. We could further apply such resource configuration caching across multiple queries in a query workload. We describe our caching mechanism below.

For each cost model (e.g., SMJ, BHJ) and sub-plan (e.g., join operator, scan operator), we maintain an in-memory index of data characteristic keys, each of which point to the best resource configuration for those data characteristics (and sub-plan, cost model). Our current prototype keeps a sorted array of keys, with automatic resizing whenever the array gets full, and we perform a binary search for lookup. We could also layout the array as a CSB$^{+}$-Tree~\cite{csb+} for larger workloads. For each resource planning call, we first check the cache and return the resource configuration from the cache in case of a hit. In case of a miss, we run the hill climbing as described above and insert the newly found resource configuration into the cache.
We provide three types of cache lookup:

\begin{itemize}
\item \textit{Exact match.} returns a hit only when exact same data characteristics match.
\item \textit{Nearest neighbor.} returns the resource configuration corresponding to the nearest data characteristic match (within a threshold).
\item \textit{Weighted Average.} returns the weighed average of neighboring resource configurations when their data characteristics are within a threshold. 
\end{itemize}

Resource plan cache can significantly reduce the resource planning overhead when the computed resource configurations are going to be very similar anyways. We will show this in Section~\ref{sec:qroEvalOpt}.

\subsection{Query \& Resource Planning}

Finally, we discuss how we integrate our resource planner with existing query planners. 
Due to the fact that we compute the resource configurations locally for each operator, we only need to invoke the resource planner when computing the costs of a sub-plan.
Thus, we extended the \texttt{getPlanCost} method of our cost model to first perform the resource planning (or lookup in the cache) and then return the sub-plan cost.
With this, as the query planner considers different candidate sub-plans, the resource planner considers the resource space for each of them.
This makes resource planning nicely integrated, and yet easily pluggable, with the query planning.
We validated that by integrating our resource planning both with the traditional Selinger optimizer~\cite{selingerOpt} as well as a newer randomized multi-objective optimizer~\cite{fastMOQO}.


\section{Evaluation}
\label{sec:eval}

We present an evaluation of our cost-based RAQO. Since we already saw the query performance results in Section~\ref{sec:evidence}, our focus here is to evaluate the planner performance.

\textbf{Setup.} Our evaluation considers the TPC-H schema as well as a randomly generated schema that could be scaled arbitrarily. 
For TPC-H, we used the same tables and the same join edges and join selectivities (we call this the join graph) as specified in the benchmark.
For the randomly generated schema, we generate a random number of tables, each of which have a randomly picked row size between 100 and 200 bytes, and a randomly picked number of rows between 100K and 2M. We then randomly generate join edges to create the join graph (with similar join selectivities as in the TPC-H schema).
The queries consist of a set of relations that need to be joined. 
For TPC-H, we consider Q12 (single join), Q3 (two joins), Q2 (three joins), and All (joining all tables).
For randomly generated schema, we generate queries having increasing number of joins, up to as many as the number of tables.
For resource configurations, we consider a cluster of $100$ containers each having a maximum size of of $10GB$. Minimum allocation is 1 container of size $1GB$ and resources could be increased in discrete intervals of 1 on either axis.

\textbf{Testbed.} We ran our evaluation on a MacBook Pro running macOS Sierra, having 2.9 GHz i5 processor, 8GB memory, and 250GB of flash storage.
Each evaluation run emits the final query plan and resource configuration found, the costs associated with these plans, the total wall-clock time taken, and the total number of resource configurations explored. We ran each query $3$ times and report the average.
Unless specified otherwise, we always cleared the resource plan cache before each query run.

Our goals for the evaluation are three fold: (i)~to evaluate resource and query planning on TPC-H schema, (ii)~to evaluate the overheads of resource planning, and (iii)~to evaluate the scalability of RAQO over larger schemas and resource configurations. We present each of these below.

\subsection{RAQO Planning}

\begin{figure}[!t]
   \centering
    \includegraphics[height=0.875\columnwidth/2]{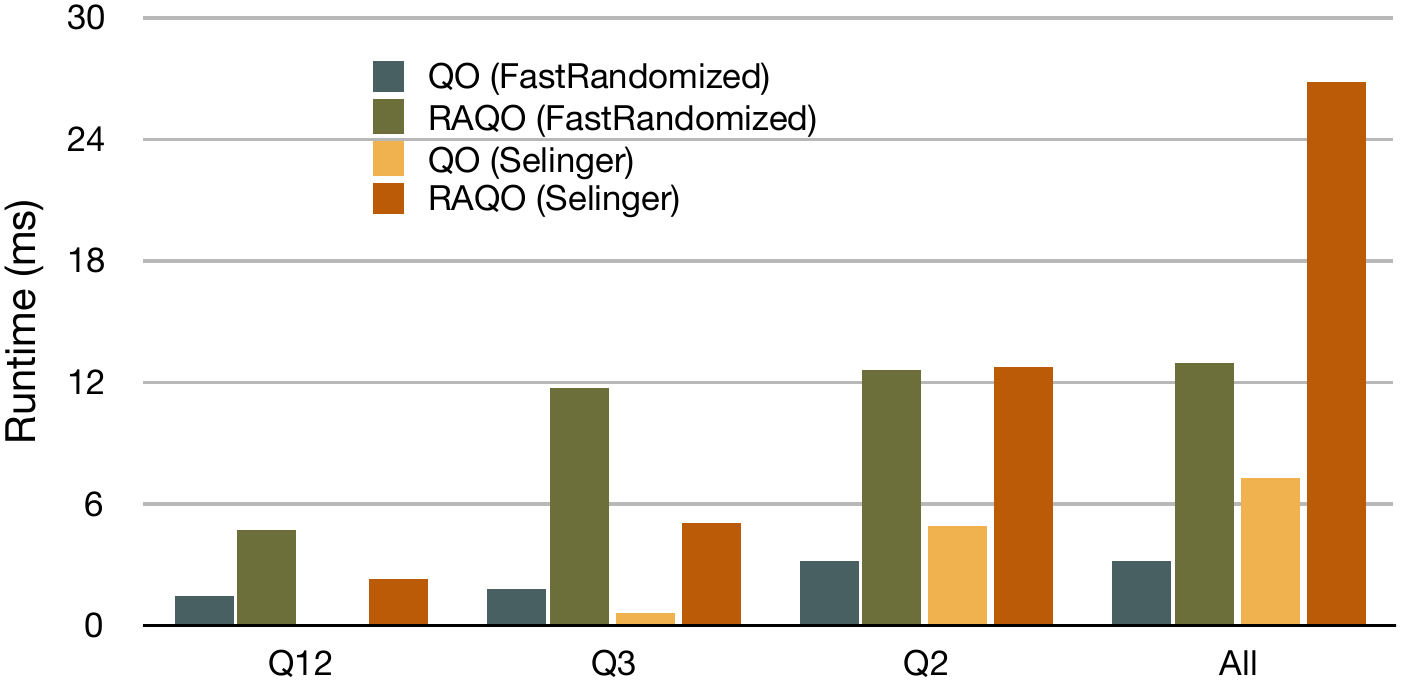}
    \caption{RAQO planning on TPC-H schema.}
    \label{fig:plannner_tpch}
\end{figure}

We tested RAQO using two query planner prototypes: 
a modern randomized algorithm to pick the best join ordering~\cite{fastMOQO}, and
a traditional System R style bottom-up join ordering algorithm (also known as Selinger optimizer)~\cite{selingerOpt}. 
We re-implemented the fast randomized algorithm as illustrated in~\cite{fastMOQO}, we refer this as \texttt{FastRandomized}.
We set the same target approximation precision as mentioned in the paper.
For each node in the plan tree, we considered the associativity and the exchange mutations as described in~\cite{joinOrderingSurvey}.
We considered two join operator implementations (SMJ and BHJ) and one scan implementation (full scan).
We ran all query planning for a default of $10$ iterations.
For System R style optimization, we implemented the Selinger algorithm for left deep trees and used the same set of mutations and operator implementations as above.
We refer this as \texttt{Selinger}.

Figure~\ref{fig:plannner_tpch} shows the planner runtimes on the TPC-H schemas. The RAQO versions of the planner ran with hill climbing but without resource plan caching. We can see that we could still generate both the resource \textit{and} the query plans in a few milliseconds. However, resource planning does add an overhead to the standard query planning. This is because of the large resource space being explored per-operator and per-candidate query plan. To illustrate, the \texttt{FastRandomized} planner considers more than half a million resource configurations for the TPC-H All query. This number goes up to over a million for the \texttt{Selinger} approach.

Below we further analyze this resource planning overhead.

\subsection{Resource Planning Overhead}
\label{sec:qroEvalOpt}

\begin{figure}[!t]
    \hspace{-0.35cm}
    \subfigure[Iterations]{
    \includegraphics[height=0.85\columnwidth/2]{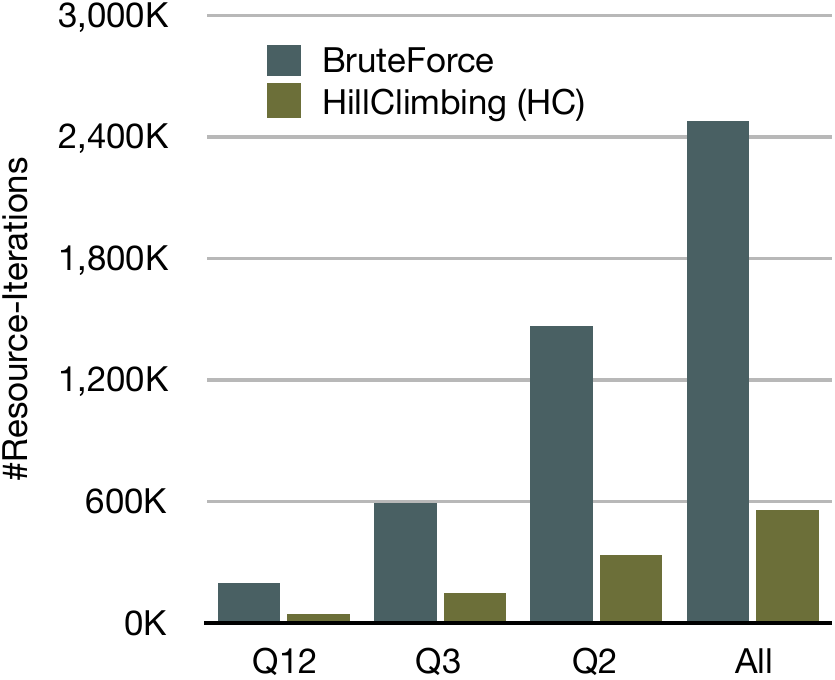}
    \label{fig:plannner_hillClimbing_iterations}
    }
    \hspace{-0.45cm}
    \subfigure[Runtime]{
    \includegraphics[height=0.85\columnwidth/2]{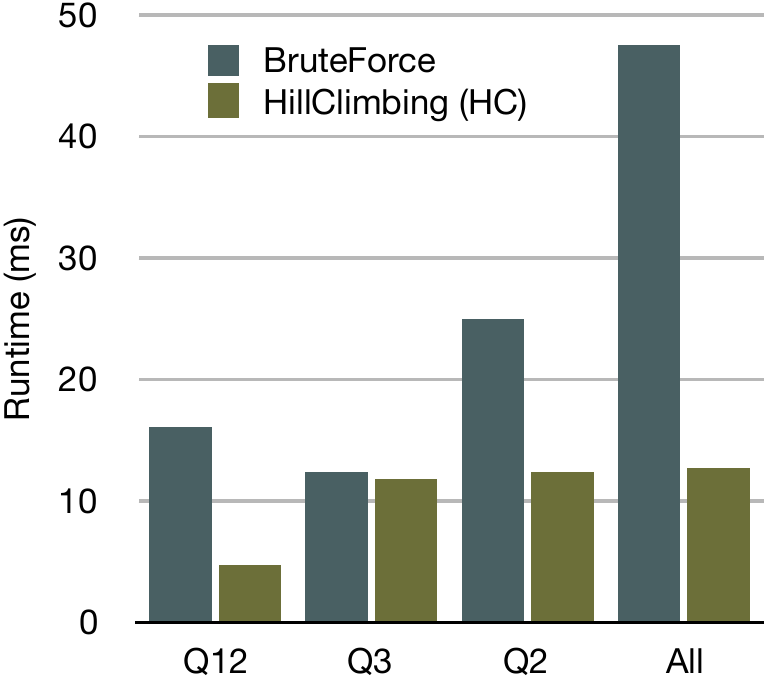}
    \label{fig:plannner_hillClimbing_runtime}
    }  
    \caption{Comparing HillClimbing with Brute Force on TPC-H schema.}
    \label{fig:planner_hillClimbing}
\end{figure}

Let us now look at the resource planning overhead. Recall that we presented two techniques for handling the large resource space: (i)~a hill climbing approach, and (ii)~resource plan caching. Below we evaluate both of these.

Figure~\ref{fig:plannner_hillClimbing_iterations} shows the number of resource configurations explored using hill climbing and brute force respectively. In general, hill climbing explores $4$ times less resource configurations than brute force. Figure~\ref{fig:plannner_hillClimbing_runtime} shows the corresponding change in runtimes. We observe similar improvements in runtime as well.

\begin{figure}[!t]
    \hspace{-0.35cm}
    \subfigure[Iterations]{
    \includegraphics[height=0.85\columnwidth/2]{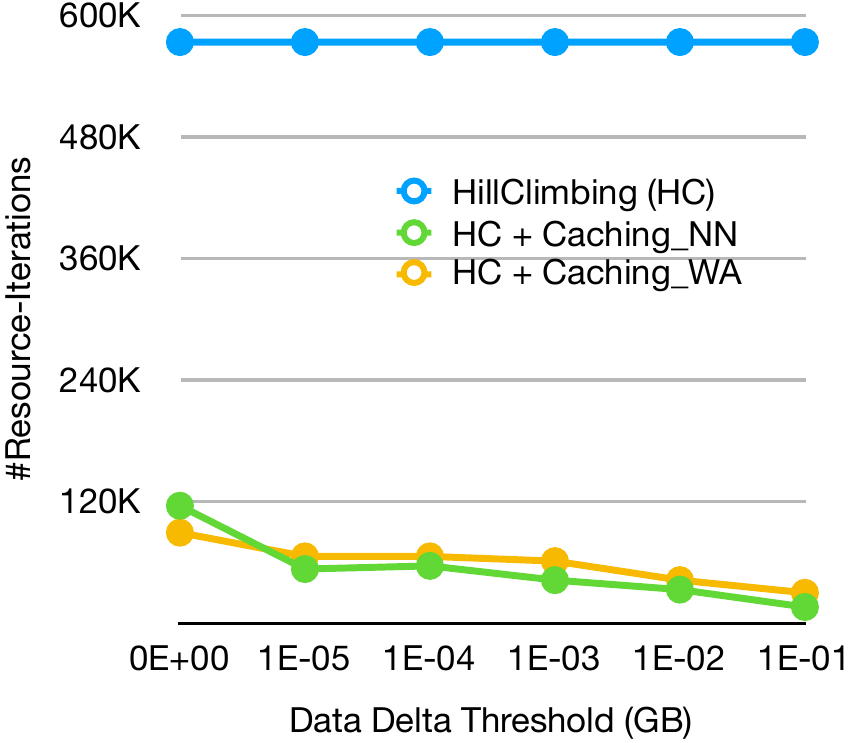}
    \label{fig:plannner_caching_iterations}
    }
    \hspace{-0.45cm}
    \subfigure[Runtime]{
    \includegraphics[height=0.85\columnwidth/2]{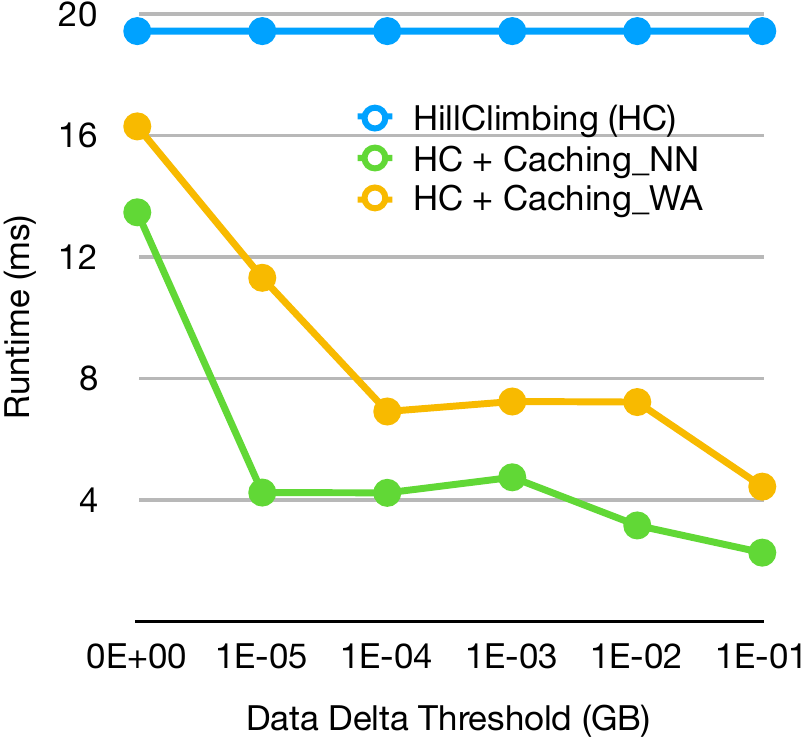}
    \label{fig:plannner_caching_runtime}
    }  
    \caption{Comparing the effectiveness of caching on TPC-H schema.}
    \label{fig:planner_caching}
\end{figure}

Figures~\ref{fig:plannner_caching_iterations} and~\ref{fig:plannner_caching_runtime} show the number of resource configurations explored and the planner runtime with and without the resource plan cache. All measurements here are on TCP-H All query.
We denote the nearest neighbor variant of our cache lookup as \texttt{HC+Caching\_NN}, while the weighted average variant is denoted as \texttt{HC+Caching\_WA}. Both variants first look for exact match before trying the interpolation. We vary the interpolation threshold, in terms of the smaller input size, on the X-axis.
We observe two things from Figures~\ref{fig:plannner_caching_iterations} and~\ref{fig:plannner_caching_runtime}: (i)~as desired, resource plan caching becomes more effective as we increase the interpolation, and (ii)~both the number of resources configurations and the planner runtime decrease significantly with resource plan caching (up to $10x$ planner time reduction for $0.1GB$ threshold).

Thus, both the hill climbing method as well as resource plan caching are effective in significantly reducing the resource planning overhead in RAQO.

\subsection{RAQO Scalability}

\begin{figure}[!t]
    \hspace{-0.35cm}
    \subfigure[Scaling Schema]{
    \includegraphics[height=0.85\columnwidth/2]{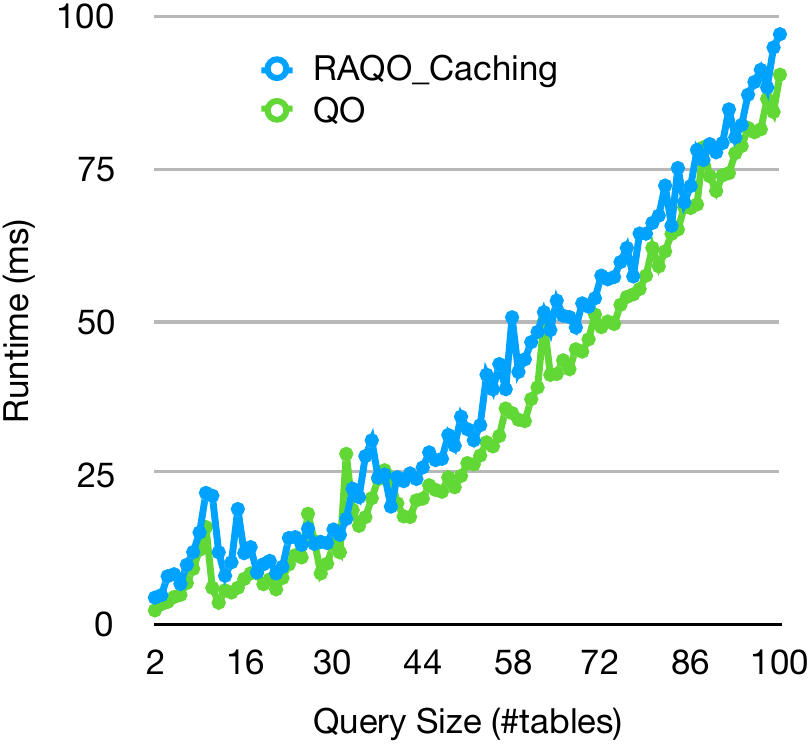}
    \label{fig:plannner_scale_schema}
    }
    \hspace{-0.45cm}
    \subfigure[Scaling Resources]{
    \includegraphics[height=0.85\columnwidth/2]{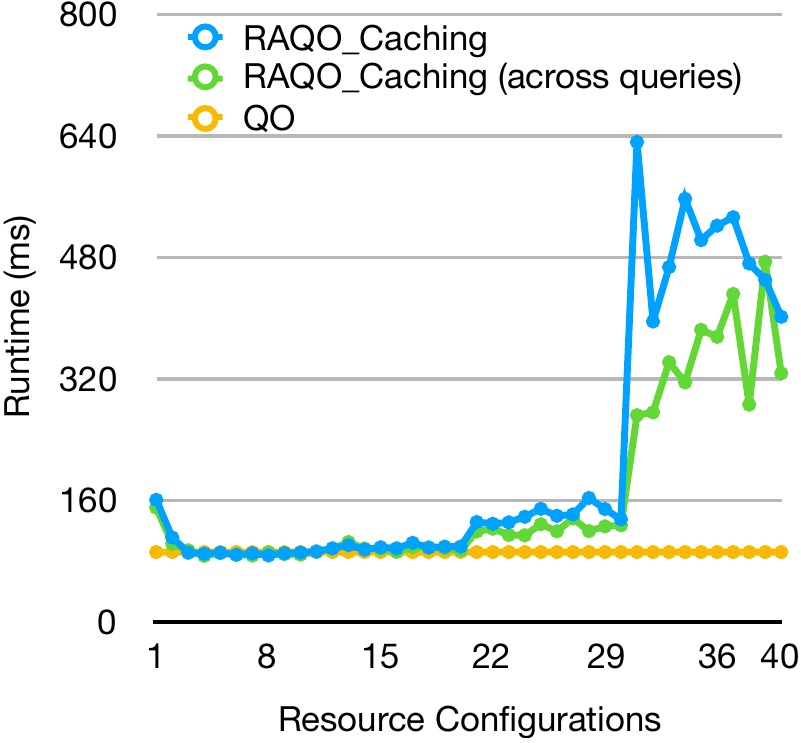}
    \label{fig:plannner_scale_resources}
    }  
    \caption{RAQO scalability over varying schema size and number of resource configurations.}
    \label{fig:planner_caching}
\end{figure}

Finally, we evaluate the scalability of the RAQO planner with larger schema sizes and larger resource configurations. We do this one at a time below.

To evaluate the scalability with schema sizes, we used the randomly generated schema (consisting of 100 tables), and ran queries with increasingly larger number of relations. 
Figure~\ref{fig:plannner_scale_schema} shows the results. The RAQO planner here uses hill climbing as well as resource plan caching.
We can see that RAQO planner performance follows closely with that of QO planner and it scales well to larger schema.
The cached version of RAQO improves over the non-cached version by almost $6x$, while it is slower than the plain QO only by a factor of $1.29x$ on average.

Next we scale the resource configuration space. For this purpose, we took the largest query in Figure~\ref{fig:plannner_scale_schema}, i.e., the one with $100$ table joins and increased the maximum cluster capacity from $100$ to $100K$ containers (in multiples of $10$) with maximum container size from $10GB$ to $100GB$ (in increments of $10GB$), giving a total of $40$ cluster conditions. Note that these numbers are realistic in modern enterprise clusters. 
Figure~\ref{fig:plannner_scale_resources} shows the planner performance over the $40$ cluster conditions.
We can see that the resource planning overhead is negligible up to $1000$ containers, it is $50\%$ for up to $10K$ containers, but the overhead climbs to $5x$ on average for more than $10K$ containers. Though the planner runtimes are still within $630$ milliseconds. 
Figure~\ref{fig:plannner_scale_resources} also shows the RAQO planner with caching turned on across queries, i.e., we do not clear the cache before each query and successive queries can leverage the older cache. Such across-query caching is indeed useful after $10K$ containers, with almost $30\%$ improvements in planner runtime.

Overall, we conclude that RAQO planner could scale well to larger schemas and to much larger cluster sizes.

\section{Research Landscape}
\label{sec:research}



After showing our first evidence about the importance of RAQO and
presenting a prototype implementation for combining query and resource optimization, we now discuss
our agenda going further, which includes several open challenges.

\vspace{-1mm}
\mypara{Explore query/resource search space}
Having to also choose the right resource configuration for each operator 
 significantly increases the already large search space that query optimizers 
have to deal with.
Hence, more efficient resource-aware plan enumeration techniques
need to be devised. Such techniques should identify and prune infeasible or
non-interesting query/resource plans early on (e.g., some operator
implementations are more efficient irrespective of resources, a broadcast
join requires one relation to fit in memory in systems like Spark,
etc.). Another open question is what should be the RAQO output: a decision
tree, a machine learning model, or analytical formulas?

\vspace{-1mm}
\mypara{RAQO on arbitrary queries}
In \autoref{sec:evidence} we focused on simple queries with one or two joins.
Next step is to generalize our approach to arbitrary operator DAGs. This is
challenging because:  (i)~with multiple operators the space of possible
resource configurations grows exponentially, (ii)~the operators may
interact, for instance, via interesting sort orders, and hence the
corresponding resource requirements may interact too, (iii)~if resources
between operators do not change, containers can be reused, creating
a trade-off between picking the best resources per operator and the
resources that would minimize the resource allocation cost.

\vspace{-1mm}
\mypara{Adaptive RAQO}
From the moment a query gets optimized until the moment its execution begins,
the condition of the cluster might change, especially in busy clusters.
In such a case, we might need to adapt/re-optimize the query, instead of 
waiting for resources to become available. Alternatively, RAQO could also 
pick plans that are more resilient to changes of cluster condition.

\vspace{-1mm}
\mypara{Interaction with DAG scheduler}
With RAQO, the submitted jobs now have precise resource requests. This
raises new questions for the scheduler in case the exact resources are not
available: should it delay the job, should it fail it, or should it
consider multiple query/resource plan alternatives and pick the most
appropriate at runtime? Moreover, should the scheduling of tasks to resources
adapt based on the selected plan (which could for instance affect the DAG's
critical path)?

\vspace{-1mm}
\mypara{Interface with resource manager (RM)}
It is crucial to define the right interface for the optimizer to talk to the
RM: a restricted API gives less opportunities for optimizations, while, at the
other extreme, exposing all the RM details to the optimizer raises security
concerns, especially in a public cloud environment. Coming up with the right interface
to learn current cluster conditions is an interesting challenge.

\vspace{-1mm}
\mypara{RAQO and pricing} 
So far we studied mostly the impact of RAQO with respect to performance and 
touched monetary costs briefly. However, there is much more to it.
For instance, it would  be interesting to see if our findings from RAQO can be used to suggest new pricing models for cloud
environments.

\vspace{-1mm}
\mypara{Beyond SQL}
In this work we focus on systems with a SQL-like interface. However,
RAQO can be applied to any system that needs to make query and resource
optimization decisions, such as streaming or machine learning systems.
Many of these lack the notion of a query optimizer in the first place, and so
building a RAQO system from scratch would be interesting.

\vspace{-1mm}
\mypara{Bridging two communities}
Overall, this is an initiative to combine efforts being done separately
by the database and the systems community. As the different layers of
modern big data systems need to increasingly collaborate with each other, so do
the corresponding communities.

\vspace{-1mm}
\mypara{Redefining the user's role}
Finally, we need to reconsider the user's role in a system that supports
RAQO.  Will the user simply provide the declarative queries and let the
system run on autopilot? Are there still control knobs she needs to handle?
What about troubleshooting and debugging? How will the ``explain'' command
look in such systems?

\section{Additional Related Work}
\label{sec:relatedwork}


In \autoref{sec:background}, we described the current landscape in big data
systems. Here we discuss some additional related work from the database
literature.


%


Classical query optimization picks the cheapest query plan for a given cost metric~\cite{KossmannSurvey2000}.
Recent works introduced multiple objectives in query optimization, e.g., execution time and monetary cost~\cite{mopqo}.
One could consider adding resources as an objective, which would lead to choosing the cheapest plan. 
This is different than our case where we choose both the plan and the resources. 
Moreover, unlike our scenario, these works do not consider dynamically changing resources.

\hide{
Classical query optimization picks the cheapest query plan for a given cost metric~\cite{KossmannSurvey2000}.
Recent works consider multiple objectives, e.g., execution time and monetary cost, using both exhaustive~\cite{mq} and approximate techniques~\cite{mq-approx}.
One might think of using resources as one of the objectives, however, we would be then essentially picking the cheapest resources.
Rather, our focus in this paper is to pick the \textit{right} set of resources for a given query. This involves picking the query plan and the resources at the same time.
}

Dynamic query evaluation plans~\cite{dynOpt94} and parametric query 
optimization~\cite{pq} 
defer some optimization decisions at runtime, by adding parameters, such as operator selectivity and table sizes, to the plans produced at compilation.
Ganguly~\cite{pq} also mentions available resources (e.g., memory size) as a parameter. However, that work does not consider the shared resource environment, where the optimizer also needs to pick the resources to be requested.

\hide{
Multi-Objective Parametric Query Optimization~\cite{mpq} generalizes parametric and multi-objective query optimization into a single optimization problem.
Essentially, picking relevant query plans for query parameters than are not known at optimization time along with multiple cost metrics.
This does not consider the dynamically changing nature of available resources and the goal is still to finally pick a query plan.
Our problem is a bit different: we look at dynamically changing resources and our goal is to pick both the query plan as well as the resources to run that plan.
}


%
%
%
%
%


\hide{
Finally, there is a rich body of work to make databases hardware-aware or hardware-conscious.
These include optimizing data layouts for better I/O performance both on disk~\cite{decompStorage,nobits} and in memory~\cite{pax,bitWeaving}, 
making better use of modern CPUs~\cite{monetdbx100},
advanced compiler techniques to generate byte codes optimized for the underlying hardware~\cite{hyperLLVM,legoBase,holger16},
or simply redesigning database systems for newer hardware~\cite{newHardwareDB,voltdb}.
These approaches, however, still consider a dedicated set of hardware resources.
}


Finally, there is a rich body of work to make databases hardware-aware or hardware-conscious.
These include optimizing data layouts for better I/O performance both on disk~\cite{nobits} and in memory~\cite{bitWeaving}, 
making better use of modern CPUs~\cite{monetdbx100},
advanced compiler techniques to generate byte codes optimized for the underlying hardware~\cite{holger16},
or simply redesigning database systems for newer hardware~\cite{voltdb}.
These approaches, however, still consider a dedicated set of hardware resources.

\section{Conclusion}

This paper opens the book for combining query and resource optimization in big data systems.
This is a major departure from current systems which treat query optimization as an upfront process, while resource management is a dynamic runtime activity.
We argue that there is a strong interplay between query plans and resource configurations, and the one cannot be chosen independently from the other.
Consequently, the query optimizer and the resource manager need to be aware of each other in order to produce much superior query plans and to avoid significant wastage of resources.
We introduce our vision of query \emph{and} resource optimization, present evidence to support our claims, describe a tool to help existing query optimizers make better decisions, and discuss an extensive research agenda.

\bibliographystyle{IEEEtran}

\bibliography{IEEEabrv,IEEEexample}

\end{document}